\documentstyle[prl,aps,twocolumn,floats,epsf]{revtex}

\begin{document}

\draft

\title{Number-phase-squeezed few-photon state generated from squeezed
atoms}

\author{Hiroki Saito and Masahito Ueda}
\address{Department of Physical Electronics,
Hiroshima University, Higashi-Hiroshima 739-8527, Japan \\
and CREST, Japan Science and Technology Corporation (JST)}


\maketitle

\begin{abstract}
This paper develops a method of manipulating the squeezed atom state to
generate a few-photon state whose phase or photon-number fluctuations are
prescribed at our disposal.
The squeezed atom state is a collective atomic state whose quantum
fluctuations in population difference or collective dipole are smaller
than those of the coherent atom state.
It is shown that the squeezed atom state can be generated by the
interaction of atoms with a coherent state of the electromagnetic field,
and that it can be used as a tunable source of squeezed radiation.
A variety of squeezed states, including the photon-number squeezed state
and the phase squeezed state, can be produced by manipulating the atomic
state.
This is owing to the fact that quantum-statistical information of the
atomic state is faithfully transferred to that of the photon state.
Possible experimental situations to implement our theory are discussed.
\end{abstract}


\narrowtext

\section{Introduction}

Nonclassical properties and manipulation of the quantized electromagnetic
(EM) field have been a center of interest in quantum
optics~\cite{squeeze}.
A variety of methods have been proposed for the generation of squeezed
states of the EM field, and several of them have been realized
experimentally~\cite{Kimble}.
The quadrature-amplitude squeezed state in which fluctuations of the
in-phase or out-of-phase component are suppressed to below those of the
coherent state can be generated via nonlinear optical
processes~\cite{Slusher,Wu}.
The photon-number squeezed state exhibiting sub-Poissonian photon
statistics can be generated using semiconductor
lasers~\cite{Machida87},
light-emitting diodes~\cite{Tapster,Hirano},
and tailor-made semiconductor heterostructures~\cite{Yamanishi}.

A coherent state with average photon number $\bar n$ has the relative
photon-number fluctuation of $\Delta n / \bar n = 1 / \sqrt{\bar n}$ and
the phase fluctuation of $\Delta \phi \simeq 1 / (2\sqrt{\bar n})$.
Hence, if $\bar n \gg 1$, there is practically no need of squeezing.
In a few-photon regime, however, $\Delta n$ becomes comparable to $\bar n$
and $\Delta \phi$ becomes of the order of one.
It is thus within the few-photon regime that the manipulation of quantum
fluctuations becomes crucially important.
Such quantum-controlled few-photon states might be useful, e.g., for
optical interconnections in semiconductor microstructures and
spectroscopic diagnostics in biology.

In the present paper, we develop a method of manipulating a collective
atomic state to generate quantum-controlled few-photon
states~\cite{Saito}.
Radiation from atoms has been extensively studied in quantum
optics~\cite{Allen}, e.g., superradiance~\cite{Dicke}, resonance
fluorescence~\cite{fluorescence}, photon emission in the
cavity~\cite{Meschede}, etc.
As regards nonclassical properties of radiation, it is known that resonant 
fluorescence exhibits photon anti-bunching and sub-Poissonian
photon statistics~\cite{Carmichael,Kimble77}.
It was pointed out in Ref.~\cite{Walls} that squeezing in the resonant
fluorescence is related to quantum fluctuations in the atomic state.
However, the relation between quantum fluctuations of the collective
atomic state and those of the emitted photon state has yet to be fully
explored from the standpoint of the control of few-photon states.
The aim of the present paper is to show that we can generate
quantum-controlled few-photon states by preparing the atoms in a
{\it squeezed atom state} (SAS), which is a collective state of
quantum-mechanically correlated atoms whose quantum fluctuations in
population difference or collective dipole are suppressed to below those
of the coherent atom state (CAS)~\cite{Arecchi}.
The SAS can be generated via the interaction of atoms with
a coherent state of photons in the cavity having a high quality
factor~\cite{Waka}.
It will be shown in Sec.~\ref{s:rad} that the SAS can be used as a tunable
source of squeezed radiation.
This is owing to the fact that quantum fluctuations of the atomic state
are rather faithfully transferred to those of the emitted photon state.
It will be shown that the number-phase uncertainty relation of photons can
be manipulated only if the atoms are in the SAS.

It is well known that the state of a two-level atom can be mapped onto
that of a spin $1 / 2$.
A collection of $N$ two-level atoms can be described with a system of
spins whose magnitudes are at most $N / 2$.
In particular, if all atoms are in the same pure state, the collective
atomic state can be described by a single spin $N / 2$.
The concept of squeezing in the spin or SU(2)
algebra~\cite{Walls,Wod,Macomber,Yurke,Aravind,Barnett,Agarwal,Kitagawa,Wineland} 
provides a mathematical definition of squeezed states in a system of
two-level atoms and in other systems that can be described by the spin
algebra.
Yurke {\it et al.}~\cite{Yurke} has pointed out that the Mach-Zehnder
interferometer is described by spin, and that its phase sensitivity can
reach the fundamental limit of $2 / N$ using an $N$ particle squeezed
state.
Kitagawa and Ueda~\cite{Kitagawa} showed that such a squeezed state
can be realized using the Coulomb interaction between charged particles.
Wineland {\it et al.}~\cite{Wineland} applied the SAS to Ramsey
spectroscopy and showed that its sensitivity can surpass that of
uncorrelated atoms.

This paper is organized as follows.
Section \ref{s:review} briefly reviews the interaction between two-level
atoms and photons in the cavity.
Section \ref{s:sss} defines the SAS in terms of the spin representation of
two-level atoms and discusses its physical meaning.
Section \ref{s:generation} analyzes dynamical processes to generate the
SAS.
Section~\ref{s:rad} describes how quantum-controlled radiation is
generated from squeezed atoms.
Section \ref{s:exp} discusses possible experimental schemes to implement
our theory.
Some complicated algebraic manipulations are relegated to the appendices
to avoid digressing from the main subjects.

\section{Interaction between Photons and Two-level atoms in the lossless
cavity}
\label{s:review}

The EM-field operator in a lossless cavity can be written as
\begin{equation}
\hat{\bf E}({\bf r}) = i \sum_n \sqrt{\frac{\hbar \omega_n}{2
\varepsilon_0}} \left[ {\bf f}_n({\bf r}) \hat a_n - {\bf f}_n^*({\bf r})
\hat a_n^\dagger \right],
\end{equation}
where $\hat a_n^\dagger$ and $\hat a_n$ are the creation and annihilation
operators of the EM field for the $n$th mode, and ${\bf f}_n({\bf r})$ is
the corresponding orthonormal mode function satisfying $(\nabla^2 +
\omega_n^2 / c^2) {\bf f}_n = {\bf 0}$, $\nabla \cdot {\bf f}_n({\bf r}) =
0$, and on the boundary the tangential component is required to vanish:
${\bf f}_{n \|}({\bf r}) = {\bf 0}$.
The Hamiltonian of the EM field in the cavity is given by
\begin{equation}
\hat H_F = \sum_n \hbar \omega_n \hat a_n^\dagger \hat a_n.
\end{equation}
Here and henceforth, zero-point energies are ignored because they do not
affect the following discussions.

Suppose that atoms have the upper energy band $|e_{j \alpha} \rangle$ and
the lower energy band $|g_{j \beta} \rangle$, where $j$ distinguishes
atoms, and $\alpha$ and $\beta$ denote Zeeman sublevels, if any, of the
upper and lower energy bands, respectively.
When the sublevels in each band are degenerate, the Hamiltonian of $N$
identical atoms have the form
\begin{equation}
\hat H_A = \sum_{j = 1}^N \frac{\hbar \omega_A}{2}
\left( \sum_{\alpha} |e_{j \alpha} \rangle \langle e_{j \alpha} | -
\sum_{\beta} |g_{j \beta} \rangle \langle g_{j \beta} | \right),
\end{equation}
where $\hbar \omega_A$ is the energy separation between the two bands.
We consider a situation in which a collection of two-level atoms is
placed in the cavity, and interacts with the EM field via the
electric-dipole interaction described by
\begin{equation} \label{dipoleH}
\hat H_I = -\sum_{j = 1}^N \hat{{\bf D}}_j \cdot \hat{{\bf E}}
(\hat{{\bf R}}_j),
\end{equation}
where $\hat{{\bf D}}_j = -e \sum_k (\hat{{\bf r}}_{jk} -
\hat{{\bf R}}_j)$ denotes the electric-dipole operator of the $j$th atom,
which is the sum of differences between the position of the nucleus
$\hat{{\bf R}}_j$ and the positions of the electrons $\hat{\bf r}_{jk}$
that belong to the $j$th atoms.
We neglect the dynamics of the center-of-mass motion of atoms, and replace
$\hat{{\bf R}}_j$ with a c-number.
Making the rotating-wave approximation in the Hamiltonian (\ref{dipoleH}),
we obtain
\begin{eqnarray} \label{HI}
\hat H_I & = & -\sum_{j = 1}^N \sum_{\alpha, \beta} \sum_n i \sqrt{\frac{\hbar 
\omega_n}{2 \varepsilon_0}} \nonumber \\
& & \times
\left[ {\bf f}_n({\bf R}_j) | e_{j \alpha}
\rangle \langle e_{j \alpha} | \hat{{\bf D}}_j | g_{j \beta} \rangle
\langle g_{j \beta} | \hat a_n - {\rm H.c.} \right],
\end{eqnarray}
where H.c. denotes the Hermite conjugate of the preceding term.

We assume that only a single mode of the EM field having energy $\hbar
\omega_F$ and a single state for each atomic energy band $|e_j \rangle$ and 
$|g_j \rangle$ participate in the interaction, and omit the subscripts
$n$, $\alpha$, and $\beta$ in the following discussions.
The Hamiltonian (\ref{HI}) then reduces to
\begin{equation} \label{HI2}
\hat H_I = \sum_{j = 1}^N \frac{1}{2} \left[ {\bf d}_j \cdot \bbox{\cal
E}({\bf R}_j) \; \hat a \hat s_{j+} + {\bf d}_j^* \cdot \bbox{\cal
E}^*({\bf R}_j) \hat a^\dagger \hat s_{j-} \right],
\end{equation}
where $\bbox{\cal E}({\bf R}_j) = -i \sqrt{2 \hbar \omega / \varepsilon_0}
{\bf f}({\bf R}_j)$ is the amplitude of the electric field per photon,
${\bf d}_j = \langle e_j | \hat{{\bf D}}_j | g_j \rangle$ is the
electric-dipole matrix element, and $\hat s_{j+} \equiv |e_j \rangle
\langle g_j |$ and $\hat s_{j-} \equiv |g_j \rangle \langle e_j |$ are the
raising and lowering operators for the $j$th atom.
Provided that the dipole moment is the same for all atoms, the
subscript $j$ in ${\bf d}_j$ may be omitted.
We define three operators $\hat s_{jx} \equiv (\hat s_{j+} + \hat s_{j-})
/ 2$, $\hat s_{jy} \equiv (\hat s_{j+} - \hat s_{j-}) / 2i$, and $\hat
s_{jz} \equiv (|e_j \rangle \langle e_j | - |g_j \rangle \langle g_j |) /
2$, which can be verified to obey the spin commutation relation $[\hat
s_{jx}, \hat s_{j'y}] = i \delta_{jj'}  \hat s_{jz}$ and its cyclic
permutations.
The two-level atom can therefore be described by spin $1/2$.
The subscripts $x$, $y$, and $z$ do not denote spatial directions, but the
expectation value of the operator $\hat s_{jz} + \frac{1}{2}$ represents
the probability of the $j$th atom being found in the excited state, and
$\hat s_{jx}$ and $\hat s_{jy}$ indicate the quadrature-phase components
of the oscillating dipole.
This can be seen by rewriting the dipole operator in the form
\begin{eqnarray}
\hat{{\bf D}}_j & = & {\bf d} | e_j \rangle \langle g_j | +
{\bf d}^* | g_j \rangle \langle e_j | \nonumber \\
& = & {\bf d} \hat s_{j+} + {\bf d}^* \hat s_{j-}
\nonumber \\
& = & 2 \left[ {\rm Re}({\bf d}) \hat s_{jx} - {\rm Im}({\bf d}) \hat
s_{jy} \right].
\end{eqnarray}
The spatial direction of the dipole depends on how we excite atoms.
For example, if the electric field at the position of an atom is linearly
polarized, the dipole oscillates along the same direction.
If the electric field at the position of the atom is circularly polarized,
the dipole also rotates in time.

Suppose that all atoms are located in a region small in comparison with
the wavelength of the field, but that they are not located too closely
together in order to avoid direct interactions between them.
The Hamiltonian of the entire system is then given by
\begin{equation} \label{JCH}
\hat H = \hbar \omega_F \hat a^\dagger \hat a + \hbar \omega_A \hat S_z +
\hbar g \left( \hat a \hat S_+ + \hat a^\dagger \hat S_- \right),
\end{equation}
where the coupling constant $g \equiv \bbox{\cal E}({\bf R}_j) \cdot
{\bf d} / (2 \hbar)$ is taken to be real without loss of generality,
and the collective spin operators are defined by
\begin{equation} \label{collectiveS}
\hat S_{\mu} \equiv \sum_j \hat s_{j \mu} \;\;\; (\mu = x, y, z), 
\end{equation}
and $\hat S_\pm \equiv \hat S_x \pm i\hat S_y$.
It is easy to show that these collective operators follow the commutation
relation of spin, $[\hat S_x, \hat S_y] = i \hat S_z$, and its cyclic
permutations. 
The Hamiltonian (\ref{JCH}) is referred to as the Jaynes-Cummings (JC)
Hamiltonian~\cite{JC}.

It is worth pointing out that one can introduce the collective spin
operators when the magnitudes of $\bbox{\cal E}({\bf R}_j) \cdot
{\bf d}_j$ in the Hamiltonian (\ref{HI2}) are the same for all the atoms
but their phases are different due, e.g., to different spatial locations
of the atoms.
The collective spin operators in this case may be defined as
\begin{equation} \label{sprime}
\hat S_\pm' \equiv \sum_j e^{\pm i\phi_j} \hat s_{j \pm}, \;\;
\hat S_z \equiv \sum_j \hat s_{j z},
\end{equation}
where $\phi_j$ is the phase of $\bbox{\cal E}({\bf R}_j) \cdot {\bf
d}_j$.
For example, when atoms are located in a one-dimensional standing wave at
every half wavelength, we have $\phi_j = j \pi$.
The operators (\ref{sprime}) also satisfy the spin commutation relations
and the Hamiltonian of the system is given by Eq.~(\ref{JCH}) in which
$\hat S_\pm$ is replaced by $\hat S_\pm'$.
Even if the spin state described by the operators (\ref{sprime}) and that
described by the operators (\ref{collectiveS}) are the same, the
corresponding states of atoms are different.
When atoms are located in the same place the dipoles oscillate in phase.
When they are located at every half wavelength, the neighboring dipoles
oscillate out of phase.
Nevertheless, the photon states generated by these atoms via the JC
Hamiltonian (\ref{JCH}) are the same.

When we move to the rotating frame for both the photon field and the atoms 
via a unitary transformation $\hat U_0(t) = e^{i (\omega_F \hat a^\dagger
\hat a + \omega_A \hat S_z) t}$, the Hamiltonian (\ref{JCH}) is
transformed to
\begin{equation} \label{Htransform}
\hat U_0 \hat H \hat U_0^\dagger + i \hbar \frac{\partial \hat
U_0}{\partial t} \hat U_0^\dagger = g \hbar (\hat a \hat S_+ e^{-i\delta
t} + \hat a^\dagger \hat S_- e^{i\delta t}),
\end{equation}
where $\delta = \omega_F - \omega_A$ denotes the detuning between the
atoms and the field.
When $\delta$ is zero, Eq.~(\ref{Htransform}) becomes
\begin{equation} \label{H}
\hat H^{\rm rot} = \hbar g \left( \hat a \hat S_+ + \hat a^\dagger \hat S_- \right).
\end{equation}
This commutes with the rotation operator,
\begin{equation} \label{rotope}
\hat U(\varphi) \equiv e^{-i \varphi (\hat a^\dagger \hat a + \hat S_z)},
\end{equation}
and is therefore invariant under rotation.
This rotational invariance allows us to choose a convenient frame of
reference without loss of generality.
For instance, when initially the EM field is in the coherent state
$|\alpha \rangle$ and the atoms are in the fully excited state $|S, M = S
\rangle$, we can arbitrarily choose the phase of the initial coherent
state without loss of generality.
Time development from the other initial state $|\alpha e^{-i \varphi}
\rangle |S, M = S \rangle$ can be obtained by a mere rotation $\hat
U(\varphi)$.

\section{Squeezing in collective two-level atoms}
\label{s:sss}

As shown in the preceding section, a collection of two-level atoms can be
described by collective spin operators (\ref{collectiveS}).
An eigenvalue of the Casimir operator $\hat{{\bf S}}^2 = \hat S_x^2 +
\hat S_y^2 + \hat S_z^2$ is given by $S (S + 1)$, where the total spin $S$
can take on values, $S = \frac{N}{2}$, $\frac{N}{2} - 1$, $\cdots$, 0 (or
$1 / 2$) when the number of atoms $N$ is even (or odd).
For each total spin $S$ there are $N! (2 S + 1) / [(\frac{1}{2}N + S + 1)!
(\frac{1}{2} N - S)!]$ different subspaces.
Generally speaking, a state of $N$ two-level atoms can be described by a
mixture of these subspaces.

Because the JC Hamiltonian (\ref{JCH}) is described by the collective spin
operators which never mix the subspaces having different total spins, we
will restrict our discussions to a single subspace having the maximal
total spin $N / 2$.
This state can be most easily accessed from the state in which all the
atoms are either in the ground state or in the excited state.
It is interesting to note that the subspaces having the same total spin 
behave exactly the same within the JC model if the numbers of atoms are
different.
For example, the state of two atoms having the total spin 1 and that of
100 atoms having the same total spin 1 obey the same JC Hamiltonian.
No single-mode photon field distinguishes between these atomic states
through the JC interaction.

A state of the single-mode photon field is defined as squeezed if, for a
nonzero range of parameter $\phi$, $\langle (\Delta \hat a_\phi)^2 \rangle$
is smaller than that of the coherent state --- the standard quantum limit
(SQL) --- of $1 / 4$, where $\hat a_\phi$ is defined as
\begin{equation} \label{aphi}
\hat a_\phi \equiv \frac{1}{2} (\hat a e^{-i\phi} + \hat a^\dagger
e^{i\phi}).
\end{equation}
The canonical commutation relation is given by $[\hat a_\phi, \hat a_{\phi 
+ \pi / 2}^\dagger] = i / 2$, and the conventional in-phase and
out-of-phase components $\hat a_1$ and $\hat a_2$ can be expressed as
$\hat a_1 = \hat a_{\phi = 0}$ and $\hat a_2 = \hat a_{\phi = \pi / 2}$,
respectively.
From the commutation relation we have
\begin{equation} \label{ucrel}
\langle (\Delta \hat a_\phi)^2 \rangle \langle (\Delta \hat
a_{\phi + \pi / 2})^2 \rangle \geq \frac{1}{16}.
\end{equation}
The coherent state has the variance of $\langle (\Delta \hat a_\phi)^2
\rangle = 1 / 4$ for any $\phi$ and satisfies the equality in
Eq.~(\ref{ucrel}).
The profile of quantum fluctuations of a photon state described by a
density operator $\hat \rho_F$ can be visualized with the quasi-probability
distribution
\begin{equation} \label{Q}
Q(\alpha) \equiv \frac{1}{\pi} \langle \alpha | \hat \rho_F | \alpha
\rangle,
\end{equation}
where $|\alpha \rangle$ is the coherent state with amplitude $\alpha$.
The quasi-probability distribution of the coherent state is isotropic and
that of the quadrature-amplitude squeezed state is elliptic.

The coherent state of a spin-$S$ system is defined by
\begin{eqnarray} \label{CSS}
|\theta, \phi \rangle & \equiv & \exp[i \theta (\hat S_x \sin\phi -
\hat S_y \cos\phi)] |S, M = S \rangle \nonumber \\
& = & \sum_{M = -S}^S \left( \begin{array}{c} 2 S \\ S + M \end{array}
\right)^{\frac{1}{2}} e^{i(S - M) \phi} \nonumber \\
& & \times \left( \sin\frac{\theta}{2}
\right)^{S - M} \left( \cos\frac{\theta}{2} \right)^{S + M} |S, M \rangle, 
\end{eqnarray}
which is referred to as the coherent spin state (CSS) or the Bloch
state~\cite{Arecchi}.
The mean spin vector of the CSS $|\theta, \phi \rangle$ points in the
direction ${\bf n} = \langle \hat{{\bf S}} \rangle / |\langle
\hat{{\bf S}} \rangle| = (\sin\theta \cos\phi, \sin\theta \sin\phi,
\cos\theta)$, where $|\langle \hat{{\bf S}} \rangle | = (\langle \hat S_x 
\rangle^2 + \langle \hat S_y \rangle^2 + \langle \hat S_z \rangle^2)^{1 /
2}$.
Denoting $\bf m$ as the unit vector that is normal to both ${\bf n}$ and
the unit vector of the $z$ direction ${\bf e}_z$, namely, ${\bf m} = {\bf n}
\times {\bf e}_z / |{\bf n} \times {\bf e}_z| = (\sin\phi, -\cos\phi, 0)$, 
we may express $|\theta, \phi \rangle$ as
\begin{equation}
|\theta, \phi \rangle = \exp[ i \theta {\bf m} \cdot \hat{{\bf S}}] |S, M
= S \rangle.
\end{equation}
When a system of two-level atoms is described by Eq.~(\ref{CSS}) in the
spin representation, we will say that the atoms are in a coherent atom
state (CAS).
The component of $\hat{{\bf S}}$ normal to the mean spin vector is given
by
\begin{equation}
\hat S({\bf n}, \chi) = \exp(-i \chi \hat{{\bf S}} \cdot {\bf n})
(\hat{{\bf S}} \cdot {\bf m}) \exp(i \chi \hat{{\bf S}} \cdot
{\bf n}),
\end{equation}
where $\chi$ denotes the angle defined on the plane normal to the mean
spin vector (see Fig.~1(a)).
\begin{figure}[tb]
\begin{center}
\leavevmode\epsfysize=105mm \epsfbox{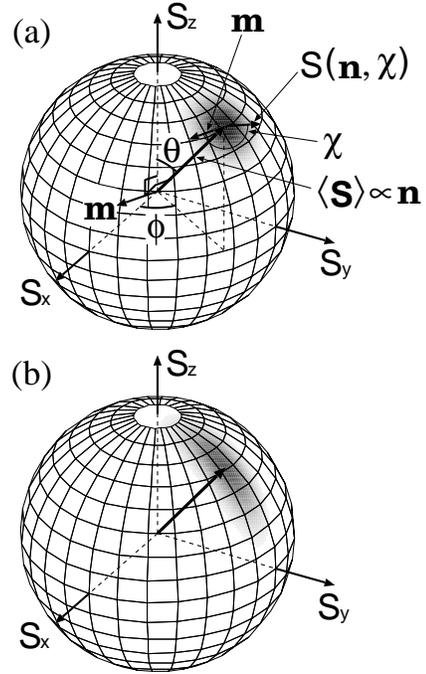}
\end{center}
\caption{
The quasi-probability distributions of (a) the coherent spin state and (b)
the squeezed spin state.
The unit vector ${\bf n}$ points in the direction of the mean spin vector,
and the unit vector ${\bf m}$ is normal to both ${\bf n}$ and the $S_z$
direction.
The spin component $S({\bf n}, \chi)$ is normal to the mean spin vector,
and the angle $\chi$ is measured from ${\bf m}$.
}
\label{f:Bloch}
\end{figure}
The commutation relation between the two quadrature components is given by 
\begin{equation}
[ \hat S({\bf n}, \chi), \hat S({\bf n}, \chi + \pi / 2)] = i
\hat{{\bf S}} \cdot {\bf n},
\end{equation}
and the corresponding uncertainty relation is given by
\begin{equation} \label{ucrelspin}
\langle [\Delta \hat S({\bf n}, \chi)]^2 \rangle \langle [\Delta
\hat S({\bf n}, \chi + \pi / 2)]^2 \rangle \geq \frac{|\langle
\hat{{\bf S}} \rangle|^2}{4}.
\end{equation}
The CSS satisfies the equality in the uncertainty relation
(\ref{ucrelspin}), and $\langle [\Delta \hat S({\bf n}, \chi)]^2
\rangle = S / 2$ for any $\chi$.
The CSS therefore has an isotropic fluctuation normal to the mean spin
vector as shown in Fig.~\ref{f:Bloch}(a), where the spin state is
visualized with the quasi-probability distribution of spin defined by
\begin{equation} \label{Qspin}
Q_s(\theta, \phi) = \frac{2 S + 1}{4\pi} \langle \theta, \phi | \hat 
\rho_A | \theta, \phi \rangle,
\end{equation}
where $\hat \rho_A$ is the density operator of a collective atomic state.
Analogous to the case of photons, a spin state is defined as squeezed if
the following inequality holds for a certain $\chi$:
\begin{equation} \label{SSSdef}
\langle [\Delta \hat S({\bf n}, \chi)]^2 \rangle < \frac{|\langle
\hat{{\bf S}} \rangle |}{2}.
\end{equation}
That is, the squeezed spin state (SSS) is a state whose fluctuation of
one component normal to the mean spin vector is less than half of the
length of the mean spin vector.
When the condition (\ref{SSSdef}) is satisfied, the variance of the
quadrature component $\langle [\Delta \hat S({\bf n}, \chi + \pi /
2)]^2 \rangle$ must be larger than $|\langle \hat{{\bf S}}
\rangle| / 2$ in order to obey the uncertainty relation (\ref{ucrelspin}),
and hence the fluctuation profile on the spin sphere becomes elliptic, as
shown in Fig.~\ref{f:Bloch}(b).

Squeezing in spin or angular momentum has been discussed by many
authors~\cite{Walls,Wod,Macomber,Yurke,Aravind,Barnett,Agarwal,Kitagawa,Wineland}.
However, the definitions of the SSS in
Refs.~\cite{Walls,Wod,Macomber,Aravind,Barnett,Agarwal} depend on the
specific spin coordinates and are therefore not invariant under rotation
in the spin space.
It was pointed out in Ref.~\cite{Kitagawa} that the direction of the
mean spin vector $\bf n$ should be taken into account to define the SSS in
a rotation-invariant manner as in Eq.~(\ref{SSSdef}).

Mathematically, SSS satisfying the condition (\ref{SSSdef}) can be
generated by unitary transformations from the CSS.
The unitary transformations have the forms $\exp(-i\xi \hat S_z^2) |\theta 
= \pi / 2, \phi \rangle$ and $\exp[-i\eta (\hat S_+^2 - \hat
S_-^2)] | \theta = 0, \phi \rangle$, where $\xi$ and $\eta$ denote the
parameters that characterize the degree of one-axis twisting and that of
two-axis countertwisting, respectively~\cite{Kitagawa}.

Let us return to the spin representation of two-level atoms.
We define the squeezed atom state (SAS) as a state of two-level atoms that
are in the SSS in the spin representation.
We note that quantum-mechanical correlations between atoms must be
established for the atoms to be in an SAS.
The state in which all atoms are in their ground state is in
a CAS $|\theta = \pi, \phi \rangle$ in the spin representation, and not
in an SAS.
If they are irradiated by a $\pi / 2$ pulse, the spin state becomes
$|\theta = \pi / 2, \phi \rangle$, which is also not squeezed, because
atoms are described by the same state and are not quantum-mechanically
correlated with each other.
We also note that a single atom cannot be squeezed, since $\langle [\Delta 
\hat S({\bf n}, \chi)]^2 \rangle$ is always $1 / 4 (= S / 2)$ for spin $1
/ 2$.
In other words, the single atom cannot be squeezed because it has no
partner with which to be quantum-mechanically correlated.

According to the definitions of the collective spin operators
(\ref{collectiveS}), $\hat S_z$ represents the population difference of
two-level atoms, and $\hat S_x$ and $\hat S_y$ represent quadrature-phase
components of the electric dipole.
Squeezing of the $\hat S_z$ component thus means reduced
fluctuations in the population difference at the expense of the enhanced
dipole fluctuation, while squeezing of $\hat S_x$,
$\hat S_y$, or their arbitrary linear combination
\begin{equation} \label{sphi}
\hat S_\phi \equiv \frac{1}{2} \left( \hat S_+ e^{-i\phi} + \hat S_-
e^{i\phi} \right).
\end{equation}
means reduced dipole fluctuations at the expense of the enhanced
fluctuations in the population difference.

To measure the $\hat S_z$ component, one can use an ionization
detector which counts the number of atoms in the excited state.
If such measurement is carried out repeatedly, with the atoms prepared in
the same state for every measurement, the variance of the population
difference $\langle (\Delta \hat S_z)^2 \rangle$ is obtained.
Variances of the other spin components can be measured by rotating the
spin state so that they become the $\hat S_z$ component.
The rotation in the spin space can be realized by irradiation of maser or
laser with classical intensity to the atoms.
The frequency of the maser or laser is assumed to be resonant with the
transition frequency of the atom.
The Hamiltonian describing the irradiation process of the classical field
is obtained by replacing the operator $\hat a$ with a c-number $\alpha$ in
the JC Hamiltonian (\ref{H}),
\begin{eqnarray} \label{Hcl}
\hat H_{\rm cl} & = & \hbar g (\alpha \hat S_+ + \alpha^* \hat S_-)
\nonumber \\
& = & 2 \hbar g |\alpha| (\hat S_x \cos\phi_c - \hat S_y \sin\phi_c),
\end{eqnarray}
where $\phi_c = \arg \alpha$ is the phase of the classical field.
The Hamiltonian (\ref{Hcl}) rotates the spin vector about the axis $\hat
S_{\phi_c}$ through angle $2g |\alpha| T_i$, where $T_i$ is the
irradiation time.
For example, the $\hat S_x$ component can be measured by counting the
population difference with the ionization detector after irradiation of
the classical field corresponding to the operation $\exp(-i\frac{\pi}{2}
\hat S_y)$.
In this operation the collective dipole of the atoms $\hat S_x$ is
converted to the population difference $\hat S_z$.

\section{Preparation of squeezed atom states}
\label{s:generation}

Several schemes for generating the SAS have been proposed.
Barnett and Dupertuis~\cite{Barnett} considered the interaction of the
antisymmetric collective dipole with the coherent EM field.
Agarwal and Puri~\cite{Agarwal} examined the steady state of atoms
interacting with broadband squeezed radiation.
Although a coordinate-dependent definition of spin squeezing $\langle
(\Delta \hat S_{x (y)})^2 \rangle < |\langle \hat S_z \rangle| / 2$ is
used in Refs.~\cite{Barnett} and \cite{Agarwal}, the states constructed
there also satisfy the coordinate-independent condition (\ref{SSSdef}).
Wineland {\it et al.}~\cite{Wineland} considered the stimulated Raman
coupling between kinetic motion of atoms in an ion trap and internal
levels of atoms, and showed that by initially squeezing the kinetic motion
one can generate the SAS of the internal state via the JC interaction.
They also showed that the coherent state of the kinetic motion can
generate the SAS via the parametric-type interaction.
Kuzmich {\it et al.}~\cite{Kuzmich} considered V-type three-level atoms
driven by squeezed light that leads to the SAS.

In the present paper we follow the scheme proposed in Ref.~\cite{Waka},
namely the interaction between the atoms and the coherent state of photons
in a high-Q cavity.
The higher-order interaction between atoms and photons establishes the
quantum correlation between the atoms, thereby reducing the dipole
fluctuation.
This scheme is simple in that no special field state, other than the
coherent state, is required.

\subsection{Analysis for the case of two atoms}
\label{s:two}

The JC model can be solved exactly for up to three atoms, and in the
zero-detuning case for up to eight atoms.
We will henceforth assume zero detuning $\delta = 0$, and employ the
Hamiltonian (\ref{H}).
By exactly solving the dynamical evolution for two atoms, we discuss the
properties of this system.

We consider the case in which both atoms are initially in the excited
state $|S = 1, M = 1 \rangle \equiv |1, 1 \rangle_A$ and photons are
in an arbitrary superposition state $\sum_n c_n |n \rangle_F$, where $|n
\rangle_F$ is the photon-number state.
The time development is calculated to be~\cite{Waka}
\begin{eqnarray} \label{atom2sol}
|\psi(t) \rangle & = & e^{-\frac{i}{\hbar} \hat H^{\rm rot} t} | n
\rangle_F |1, 1 \rangle_A \nonumber \\
& = & \sum_{n = 0}^\infty c_n e^{-i(n + 1) \omega_F t} \bigl[
 p_n(t) |1, 1 \rangle_A |n \rangle_F \nonumber \\
& & + q_n(t) |1, 0 \rangle_A |n + 1
\rangle_F + r_n(t) |1, -1 \rangle_A |n + 2 \rangle_F \bigr], \nonumber \\
& & 
\end{eqnarray}
where
\begin{mathletters}
\begin{eqnarray}
p_n(t) & = & \frac{(n + 1) \cos \sqrt{2 (2 n + 3)} gt + n + 2}{2 n + 3},
\\
q_n(t) & = & -i \sqrt{\frac{n + 1}{2 n + 3}} \sin \sqrt{2 (2 n + 3)} gt, \\
r_n(t) & = & \frac{\sqrt{(n + 1) (n + 2)}}{2 n + 3} \left( \cos \sqrt{2 (2
n + 3)} gt - 1 \right).
\end{eqnarray}
\end{mathletters}
One can calculate any physical quantities from this solution.

Let us first consider the photon-number state $|n \rangle_F$ as the
initial state.
In this case the initial state $|n \rangle_F |1, 1 \rangle_A$ is invariant
with respect to rotation (\ref{rotope}), and consequently $\langle \hat
S_x \rangle = \langle \hat S_y \rangle = 0$, which remains true at later
times.
The variances of the components normal to the mean spin vector are
calculated to be
\begin{eqnarray}
\langle (\Delta \hat S_x)^2 \rangle & = & \langle (\Delta \hat S_y)^2
\rangle \nonumber \\
& = & \frac{1}{2} \left( 1 + \frac{n + 1}{2 n + 3} \sin^2 \sqrt{2 (2 n + 3)}
gt \right),
\end{eqnarray}
which is always greater than $S / 2 = 1 / 2$, and hence the spin state can 
never be squeezed.
Generally, when the initial state is invariant with respect to the
rotation $\hat U(\varphi)$, the atoms can never be squeezed for any number
of atoms.

\begin{figure}[tb]
\leavevmode\epsfysize=74mm \epsfbox{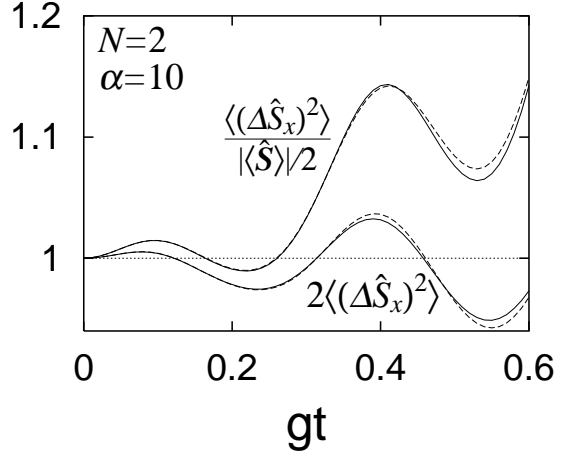}
\caption{
Time evolutions of the normalized variance $2 \langle (\Delta \hat
S_x)^2 \rangle$ and the squeezing factor $2 \langle (\Delta \hat S_x)^2
\rangle / |\langle \hat{\bf S} \rangle|$ for two atoms.
The two atoms are initially excited, and the EM field is initially in the
coherent state with amplitude $\alpha = 10$.
The solid curves show the numerical results, and the dashed ones show
approximate solutions (\protect\ref{appsx}) and (\protect\ref{sp}).
}
\label{f:twoatom}
\end{figure}

When the photon field is initially in the coherent state $|\alpha
\rangle$, the coefficients are given by $c_n = e^{-|\alpha|^2 / 2}
\alpha^n / \sqrt{n!}$.
The amplitude $\alpha$ can be taken to be real without loss of generality, 
and in this case $\langle \hat a_2 \rangle$ and $\langle \hat S_x
\rangle$ vanish at any time (see appendix \ref{s:vanish}).
Therefore the $S_x$ direction is always normal to the mean spin vector.
The variance of $\hat S_x$ is calculated to be
\begin{eqnarray} \label{atom2sx}
\langle (\Delta \hat S_x)^2 \rangle & = & e^{-|\alpha|^2} \sum_{n = 2}^\infty 
\frac{\alpha^{2 n - 2}}{\sqrt{n! (n - 2)!}} p_n(t) r_{n - 2}(t) \nonumber
\\
& & +
\frac{1}{2} e^{-|\alpha|^2} \sum_{n = 0}^\infty \frac{\alpha^{2n}}{n!}
\left[ p_n(t)^2 + 2 q_n(t)^2 + r_n(t)^2 \right]. \nonumber \\
& & 
\end{eqnarray}
When $\alpha \gg 1$, the photon-number distribution has a narrow peak
relative to the mean photon number $\bar n$, and one can expand
(\ref{atom2sx}) with respect to $n - \bar n$.
Replacing the summations with the integrals we obtain, for $gt \lesssim 1
/ \sqrt{\bar n}$,
\begin{equation} \label{appsx}
\langle (\Delta \hat S_x)^2 \rangle \simeq \frac{1}{2} - \frac{1}{2 \bar
n} \sin^4 \sqrt{\bar n} gt + \frac{gt}{2 \sqrt{\bar n}} \sin 2\sqrt{\bar
n} gt.
\end{equation}
Similarly, $\langle \hat S_y \rangle$ and $\langle \hat S_z \rangle$ are
approximated to be
\begin{mathletters}
\begin{eqnarray}
\langle \hat S_y \rangle & \simeq & -e^{-(gt)^2 / 2} \sin 2\sqrt{\bar n}
gt - \frac{gt}{\sqrt{\bar n}} \left( \frac{3}{4} - \frac{5}{2} \sin^2
\sqrt{\bar n} gt \right) \nonumber \\
& & + \frac{1}{8 \bar n} \left( \sin 2 \sqrt{\bar n} 
gt + \sin 4 \sqrt{\bar n} gt \right),
\\
\langle \hat S_z \rangle & \simeq & e^{-(gt)^2 / 2} \cos 2\sqrt{\bar n} gt
- \frac{5 gt}{4 \sqrt{\bar n}} \sin 2 \sqrt{\bar n} gt \nonumber \\
& & + \frac{1}{4 \bar n} \sin^2 2 \sqrt{\bar n} gt.
\end{eqnarray}
\end{mathletters}
Therefore, if the squeezing factor defined by
\begin{eqnarray} \label{sp}
\frac{\langle (\Delta \hat S_x)^2 \rangle}{|\langle \hat{{\bf S}}
\rangle| / 2} & \simeq & e^{\frac{(gt)^2}{2}} - \frac{1}{\bar n} \sin^2
\sqrt{\bar n} gt + \frac{3}{8 \bar n} \sin^2 2\sqrt{\bar n} gt \nonumber \\
& & + \frac{3 gt}{2 \sqrt{\bar n}} \sin 2\sqrt{\bar n} gt
\end{eqnarray}
is less than one, the condition for the SAS (\ref{SSSdef}) is fulfilled.
Figure \ref{f:twoatom} compares the time evolution of the approximate
formula (\ref{sp}) (dashed curve) with the exact one which is numerically
calculated from (\ref{atom2sol}) (solid curve) for two atoms and for $\bar
n = \alpha^2 = 100$.
The parameter $gt$ in Fig.~\ref{f:twoatom} and all the quantities
appearing in the figures presented henceforth are dimensionless.
We find that both curves are in excellent agreement and the SAS is attained
around $gt = 0.2$.
The variance of another component that is normal to both the mean spin
vector and the $S_x$ direction never reduces to below $1 / 2$.
It can be shown numerically that the SAS never occurs after the first 
minimum around $gt = 0.2$.
Although in Fig.~\ref{f:twoatom} the second minimum of the variance
$\langle (\Delta \hat S_x)^2 \rangle$ goes below the first minimum, the
squeezing factor does not go below the first minimum because the length of
the spin vector also decreases.

Equation (\ref{sp}) shows that squeezing vanishes when the intensity of 
the coherent state is sufficiently large, $\bar n \gg 1$,
which is due to the fact that the classical field merely rotates the spin
vector.
The photon-number state cannot produce the SAS, as mentioned above.
We thus find that both wave and particle aspects of photons are necessary
for atoms to be squeezed.

\subsection{Analytic approach for the case of a large number of atoms}
\label{s:analytic}

We provide here approximate analytic expressions for the case of a large
number of atoms.
These are derived by neglecting the terms of order $1 / N$ relative to the
dominant terms in the equations of motion, which are therefore very
accurate when the number of atoms $N$ is very large.

The initial state is assumed to be the totally excited state of the atoms
$|S, M = S \rangle$ and the coherent state of the field  $|\alpha
\rangle$, where $\alpha$ is assumed to be real and hence $\langle
\hat S_x \rangle = \langle \hat a_2 \rangle = 0$.
The other averages obey the equations of motion (see
appendix~\ref{s:solution} for derivations),
\begin{mathletters}
\begin{eqnarray} \label{eom}
\frac{d \langle \hat S_y \rangle}{dt} & \simeq & -2 g \langle \hat a_1
\rangle \langle \hat S_z \rangle, \\
\frac{d \langle \hat S_z \rangle}{dt} & \simeq & 2 g \langle \hat a_1
\rangle \langle \hat S_y \rangle, \\
\frac{d \langle \hat a_1 \rangle}{dt} & \simeq & -g \langle \hat S_y
\rangle, 
\end{eqnarray}
\end{mathletters}
which become those of a pendulum, if we set
\begin{mathletters} \label{peq}
\begin{eqnarray}
\langle \hat S_y \rangle & = & \frac{N}{2} \sin\theta, \\
\langle \hat S_z \rangle & = & \frac{N}{2} \cos\theta, \\
\langle \hat a_1 \rangle & = & -\frac{1}{2g} \frac{d\theta}{dt}.
\end{eqnarray}
\end{mathletters}
The solutions of Eqs.~(\ref{eom}) can be expressed in terms of Jacobi's
elliptic functions~\cite{MathFunc}.
Solving the equations of motion for fluctuations, we obtain
\begin{mathletters} \label{SOLUTIONS}
\begin{eqnarray}
\langle (\Delta \hat a_2)^2 \rangle & = & \frac{1}{4 {\rm dn}^2(u|m)}
\left[ 1 + m E^2(u|m) \right], \\
\langle (\Delta \hat S_x)^2 \rangle & = & \frac{N}{4} \Biggl\{ m \;
\frac{{\rm sn}^2(u|m) {\rm cn}^2(u|m)}{{\rm dn}^4(u|m)} \nonumber \\
& & + \Bigl[ m \;
\frac{{\rm sn}(u|m) {\rm cn}(u|m)}{{\rm dn}^2(u|m)}
E(u|m) \nonumber \\
& & + {\rm dn}(u|m) \Bigr]^2 \Biggr\},
\end{eqnarray}
\end{mathletters}
where $u \equiv gt\sqrt{N + \alpha^2}$ and $m \equiv N / (N + \alpha^2)$.
Jacobi's elliptic functions~\cite{MathFunc} are defined by
${\rm sn}(u|m) = \sin\varphi$, ${\rm cn}(u|m) = \cos\varphi$, ${\rm
dn}(u|m) = \sqrt{1 - m \sin^2\varphi}$, where $u$ and $\varphi$ are
related by
\begin{equation}
u = \int_0^\varphi \frac{d\theta}{\sqrt{1 - m \sin^2\theta}}.
\end{equation}
The elliptic integral of the second kind is given by
$E(u|m) = \int_0^u {\rm dn}^2(u'|m) du'$.

\begin{figure}[tb]
\leavevmode\epsfysize=74mm \epsfbox{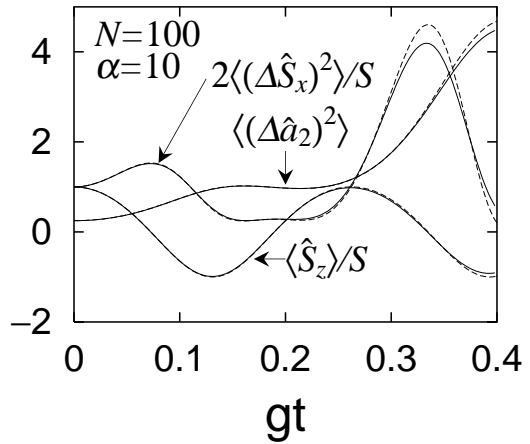}
\caption{
Time evolutions of $\langle \hat S_z \rangle / S$, $2\langle (\Delta \hat
S_x)^2 \rangle / S$ and $\langle (\Delta \hat a_2)^2 \rangle$ for 100
atoms ($S = 50$).
All the atoms are initially excited, and the EM field is initially in the
coherent state with amplitude $\alpha = 10$.
The solid curves show the numerical results, and the dashed ones show
approximate solutions (\protect\ref{SOLUTIONS}) and
(\protect\ref{sol1}b).
}
\label{f:compare}
\end{figure}

Figure \ref{f:compare} compares the analytic solutions (\ref{SOLUTIONS})
(dashed curves) with the numerically exact ones (solid curves) for 100
atoms and $\alpha = 10$.
We find that the analytic curves are in excellent agreement with the
numerical ones.
The analytic curves, however, begin to deviate from the numerical ones at
around $gt \simeq 0.3$.
This is because the differential equations (\ref{avgeom2}) and
(\ref{fleq2}) include errors of order $1 / N$ relative to the dominant
terms, which accumulate to produce errors in the solutions of order $e^{gt
\sqrt{N}} / N$, which becomes of order unity around $gt \simeq 0.3$.

The analogy to the pendulum gives us a qualitative and simple account of
the squeezing mechanism.
When the pendulum points in the direction $(\sin\theta \cos\phi,
\sin\theta \sin\phi, \cos\theta)$, it undergoes the force toward the
direction $(\cos\theta \cos\phi, \cos\theta \sin\phi, -\sin\theta)$.
In the present case, where $\alpha$ is taken to be real, the pendulum
begins to fall toward the negative $S_y$ axis and rotates on the
$S_y$-$S_z$ plane.
Suppose that the pendulum has a deviation from the $S_y$-$S_z$ plane
$(\phi = -\frac{\pi}{2} + \delta\phi)$, the direction of the force is
$(\cos\theta \delta\phi, 0, -\sin\theta)$.
This force increases the deviation when $\cos\theta > 0$, and decreases it 
when $\cos\theta < 0$.
In fact, in Fig.~\ref{f:compare}, $\langle (\Delta \hat S_x)^2 \rangle$
increases when $\langle \hat S_z \rangle > 0$, and decreases when $\langle
\hat S_z \rangle < 0$.

\subsection{Numerical analysis}
\label{s:numerical}

When the number of atoms is intermediate, analytic solutions are
unavailable, so we study the dynamical evolution of the system by
numerically diagonalizing the Hamiltonian (\ref{H}).
The amount of computation increases with increasing the number of atoms
$N$ roughly as $N^3$.
The initial state is assumed to be the totally excited state of the atoms
$|S, M = S \rangle$ and the coherent state of the photon field $|\alpha
\rangle$, where $\alpha$ is again taken to be real.

\begin{figure}[tb]
\leavevmode\epsfysize=74mm \epsfbox{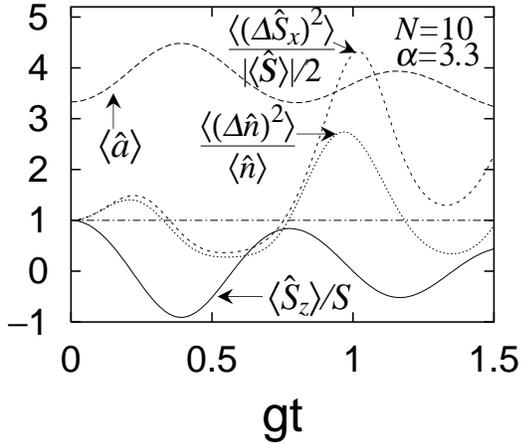}
\caption{
Time evolutions of $\langle \hat S_z \rangle / S$, $2 \langle (\Delta \hat
S_x)^2 \rangle / |\langle \hat{\bf S} \rangle|$, $\langle \hat a
\rangle$ and $\langle (\Delta \hat n)^2 \rangle / \langle \hat n \rangle$ 
for 10 atoms ($S = 5$).
All the atoms are initially excited, and the EM field is initially in the
coherent state with amplitude $\alpha = 3.3$.
}
\label{f:evolution1}
\end{figure}

\begin{figure}[tb]
\leavevmode\epsfysize=74mm \epsfbox{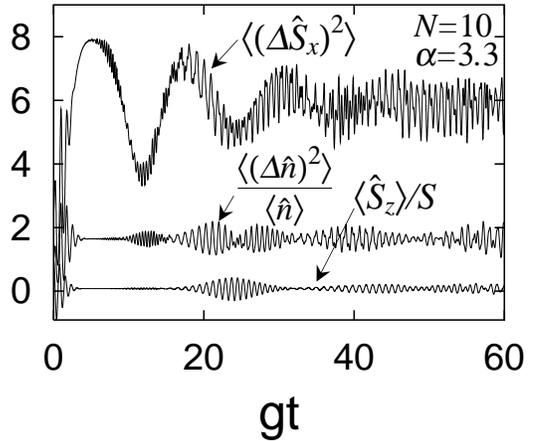}
\caption{
Long-term behaviors of $\langle \hat S_z \rangle / S$, $\langle (\Delta
\hat S_x)^2 \rangle$ and $\langle (\Delta \hat n)^2 \rangle / \langle
\hat n \rangle$ for 10 atoms.
All the atoms are initially excited, and the photon field is initially in
the coherent state with amplitude $\alpha = 3.3$.
}
\label{f:revival}
\end{figure}

Figure~\ref{f:evolution1} shows time evolutions of statistical properties
of atoms and photons.
The number of atoms is 10, and the amplitude of the initial coherent state
is chosen to be $\alpha = 3.3$ to obtain the maximal squeezing of the
atoms.
The $S_x$ component is always normal to the mean spin vector, since
$\langle \hat S_x \rangle = 0$.
In Fig.~\ref{f:evolution1}, the squeezing factor $2 \langle (\Delta \hat
S_x)^2 \rangle / |\langle \hat{{\bf S}} \rangle |$ becomes less than one, 
which indicates that the SAS is obtained.
The maximum degree of squeezing is attained in the first minimum.
It is found from the long-term behavior that the squeezing never occurs
at a later time.
The fluctuation of the other component that is normal to both the $S_x$
direction and the mean spin vector never fulfills the squeezing condition
(\ref{SSSdef}).
Since the mean spin vector rotates in the $S_y$-$S_z$ plane, $\langle \hat 
S_z \rangle$ oscillates with the amplitude of $|\langle \hat{{\bf S}}
\rangle|$.
The amplitude of the photon field also oscillates with the
same period but out of phase because of the energy exchange between the
atoms and the photon field.
The variance $\langle (\Delta \hat S_x)^2 \rangle$ increases when $\langle
\hat S_z \rangle > 0$, and decreases when $\langle \hat S_z \rangle < 0$,
as discussed in the previous subsection.
The Fano factor $\langle (\Delta \hat n)^2 \rangle / \langle \hat n
\rangle$ of the photon field also goes below the SQL, and its behavior is
very similar to that of $\langle (\Delta \hat S_x)^2 \rangle$.
The long-term behavior of this system is shown in Fig.~\ref{f:revival}.
The collapse and revival phenomena occur in the population difference and
in the Fano factor as in the case of a single atom~\cite{Eberly}.
The revival peak of the Fano factor splits and there is a
small revival before the main revival.
The variance $\langle (\Delta \hat S_x)^2 \rangle$, on the other hand,
oscillates with the same period as the revivals, and the initially regular 
oscillations gradually change to random fluctuations around some value.

\begin{figure}[tb]
\leavevmode\epsfysize=74mm \epsfbox{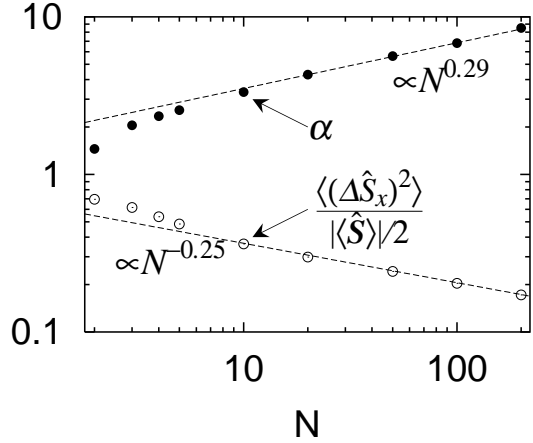}
\caption{
Minimum values of the squeezing factors $2 \langle (\Delta \hat S_x)^2
\rangle / |\langle \hat{\bf S} \rangle|$ obtained by the interaction of
atoms with the coherent states of photons as a function of the number of
atoms $N$.
For each $N$ the amplitude of the coherent state $\alpha$ is chosen to
give the best squeezing factor.
The squeezing factor tends to scale as $N^{-0.25}$ for large $N$ and the
optimal amplitude $\alpha$ as $N^{0.29}$.
}
\label{f:dependence}
\end{figure}

The degree of squeezing of the SAS depends on the number of atoms $N$, and
for each $N$ the maximum degree of squeezing is attained at a particular
amplitude $\alpha$ of the initial coherent state.
Figure \ref{f:dependence} shows the minimum squeezing factor
$2 \langle (\Delta \hat S_x)^2 \rangle / |\langle \hat{{\bf S}}
\rangle |$ for each number of atoms and the amplitude of the initial
coherent state that gives this factor.
We find that the higher degree of squeezing can be obtained for the larger
number of atoms.
The squeezing factor tends to behave as $N^{-0.25}$ when $N$ is more than
about ten, and the optimal amplitude $\alpha$ behaves as $N^{0.29}$.

\section{Quantum-controlled radiation from squeezed atoms}
\label{s:rad}

It is natural to expect that the atoms whose collective dipole or
population difference is squeezed can radiate the photon field having
nonclassical properties.
We will show that this is indeed the case, and that quantum fluctuations
of the photon field can be controlled by manipulating the SAS, which is
done by applying a classical field to the atoms.

\subsection{Radiation from squeezed atoms}

\begin{figure}[tb]
\leavevmode\epsfysize=50mm \epsfbox{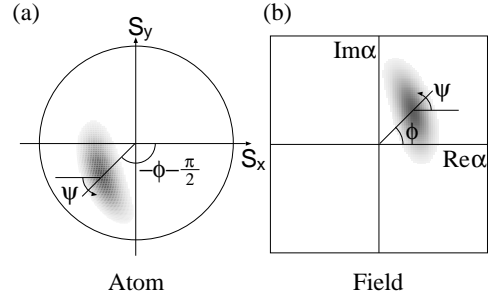}
\caption{
The relation between (a) the quasi-probability distribution of the
prepared atoms (\protect\ref{Qspin}) and (b) that of the emitted photon
field (\protect\ref{Q}).
The angle $\phi$ (or $-\phi - \pi / 2$) represents the direction of the
mean amplitude of the photon field (or the mean spin vector), and $\psi$
represents the direction of the fluctuations of the spin and the photon
field.
The $S_z$ component of the mean spin vector is negative.
}
\label{f:ellipses}
\end{figure}

The Heisenberg equations of motion for $\hat a_\phi$ and $\hat S_{-\phi
-\pi / 2}$ are written as
\begin{mathletters} \label{eomaS}
\begin{eqnarray}
\dot{\hat a}_\phi & = & \frac{i}{\hbar} [ \hat H^{\rm rot}, \hat a_\phi] = 
g \hat S_{-\phi - \pi / 2}, \\
\dot{\hat S}_{-\phi - \pi / 2} & = & \frac{i}{\hbar} [ \hat H^{\rm rot},
\hat S_{-\phi - \pi / 2}] = 2g \hat a_\phi \hat S_z.
\end{eqnarray}
\end{mathletters}
Equation (\ref{eomaS}a) indicates that the phase of the photon field is
connected with the direction of the spin vector.
When the spin vector is tilted toward the direction of $-\phi - \pi / 2$,
the field is initially amplified toward the direction of $\phi$, as
illustrated in Fig.~\ref{f:ellipses}.
The equations of motion for various fluctuations are given by
\begin{mathletters} \label{eomfluc}
\begin{eqnarray}
\frac{d}{dt} \langle (\Delta \hat a_\psi)^2 \rangle & = & 2g \langle
(\Delta \hat a_\psi) (\Delta \hat S_{-\psi - \pi / 2}) \rangle, \\
\frac{d}{dt} \langle (\Delta \hat a_\psi) (\Delta \hat S_{-\psi - \pi /
2}) \rangle & = & g \biggl[ \langle (\Delta \hat S_{-\psi - \pi / 2})^2
\rangle \nonumber \\
& & + 2 \langle (\Delta \hat a_\psi) (\Delta \hat a_\psi \hat S_z)
\rangle \biggr], \\
\frac{d}{dt} \langle (\Delta \hat S_{-\psi - \pi / 2})^2 \rangle & = & 2g 
\langle [ \Delta \hat S_{-\psi - \pi / 2}, \Delta \hat a_\psi \hat S_z ]_+ 
\rangle,
\end{eqnarray}
\end{mathletters}
where $\Delta \hat {\cal O} \equiv \hat {\cal O} - \langle \hat {\cal O}
\rangle$, and $[\hat A, \hat B]_+ \equiv \hat A \hat B + \hat B \hat A$ is
an anti-commutator.
The angle $\psi$ in Eqs.~(\ref{eomfluc}) represents the direction of the
fluctuations of the spin and the photon field, as shown in
Fig.~\ref{f:ellipses}.
The right-hand side of Eq.~(\ref{eomfluc}a) vanishes at $t = 0$, because
initially the atoms and the photon field are not correlated.
Since the first derivative vanishes at $t = 0$, the time development for
small $t$ is determined by the second derivative.
From Eqs.~(\ref{eomfluc}a) and (\ref{eomfluc}b) we have
\begin{equation} \label{da2dt}
\frac{d^2}{dt^2} \langle (\Delta \hat a_\psi)^2 \rangle = 2g^2 \left[
\langle (\Delta \hat S_{-\psi - \pi / 2})^2 \rangle + 2 \langle (\Delta
\hat a_\psi) (\Delta \hat a_\psi \hat S_z) \rangle \right].
\end{equation}
At $t = 0$, the right-hand side of Eq.~(\ref{da2dt}) reduces to $2g^2 
[\langle (\Delta \hat S_{-\psi - \pi / 2})^2 \rangle + \langle \hat S_z
\rangle / 2]$ because $\langle (\Delta \hat a_\psi)^2 \rangle = 1 /
4$ for the vacuum state.
Therefore, if the initial spin state satisfies the condition,
\begin{equation} \label{sqcond}
\langle (\Delta \hat S_{-\psi - \pi / 2})^2 \rangle < -\frac{\langle \hat
S_z \rangle}{2},
\end{equation}
the photon field will evolve into a squeezed state.
To satisfy the inequality (\ref{sqcond}), $\langle \hat S_z
\rangle$ must be negative.
The equation of motion (\ref{da2dt}) indicates that the fluctuation
profile of the photon field is connected with that of the spin state.
From Eqs.~(\ref{eomaS}a) and (\ref{da2dt}), then, the direction toward
which the spin vector tilts corresponds to the direction of the
displacement on the complex-$\alpha$ plane of the photon field, and the
squeezed or enhanced direction of the spin fluctuation corresponds to that
of the fluctuation of the photon field.
Consequently, the quasi-probability distribution of the photon field on
the complex-$\alpha$ plane is expected to behave like the
quasi-probability distribution of the atoms on the spin sphere, as
illustrated in Fig.~\ref{f:ellipses}.

When the tilting angle of the spin vector from the $z$ axis is small,
i.e., $\theta \simeq \pi$, we can approximately solve the equations of
motion (\ref{eomaS}) and (\ref{eomfluc}).
In this case, $\langle \hat S_z \rangle$ is almost constant, and $\hat
S_z$ can be replaced by a constant c-number $\langle \hat S_z \rangle_0$,
where $\langle \cdots \rangle_0$ denotes the expectation value with
respect to the initial state.
With this approximation, Eqs.~(\ref{eomaS}) can be solved, giving
\begin{mathletters} \label{ampsol}
\begin{eqnarray}
\langle \hat a_\phi \rangle & = & \frac{\langle \hat
S_{-\phi - \pi / 2} \rangle_0}{\sqrt{2 | \langle \hat S_z \rangle_0 |}}
\sin \sqrt{2 | \langle \hat S_z \rangle_0 |} gt, \\
\langle \hat S_{-\phi - \pi / 2} \rangle & = & \langle \hat
S_{-\phi - \pi / 2} \rangle_0 \cos \sqrt{2 | \langle \hat S_z \rangle_0 |}
gt.
\end{eqnarray}
\end{mathletters}
The equations of motion for the fluctuations (\ref{eomfluc}) become closed 
forms in this approximation, and the solutions are given by
\begin{mathletters} \label{flucsol}
\begin{eqnarray}
\langle (\Delta \hat a_\psi)^2 \rangle & = & \frac{1}{4} \cos^2 \sqrt{2 |
\langle \hat S_z \rangle_0 |} gt \nonumber \\
& & + \frac{\langle (\Delta \hat S_{-\psi -
\pi / 2})^2 \rangle_0}{2 | \langle \hat S_z \rangle_0 |} \sin^2 \sqrt{2 |
\langle \hat S_z \rangle_0 |} gt, \nonumber \\
& & \\
\langle (\Delta \hat S_{-\psi - \pi / 2})^2 \rangle & = & \langle (\Delta
\hat S_{-\psi - \pi / 2})^2 \rangle_0 \cos^2 \sqrt{2 | \langle \hat S_z
\rangle_0 |} gt \nonumber \\
& & + \frac{| \langle \hat S_z \rangle_0 |}{2} \sin^2 \sqrt{2
| \langle \hat S_z \rangle_0 |} gt.
\end{eqnarray}
\end{mathletters}
We find that if the condition (\ref{sqcond}) for the initial spin state is 
fulfilled, the variance of the quadrature amplitude (\ref{flucsol}a) goes
below the SQL of $1 / 4$.
At time $t = \pi (2 \sqrt{2 | \langle \hat S_z \rangle_0 |} g)^{-1}$,
the fluctuation $\langle (\Delta \hat a_\psi)^2 \rangle$ attains its first 
minimum
\begin{equation}
\langle (\Delta \hat a_\psi)^2 \rangle = \frac{\langle (\Delta \hat
S_{-\psi - \pi / 2})^2 \rangle_0}{2 | \langle \hat S_z \rangle_0 |},
\end{equation}
and at the same time the amplitude of the field becomes maximum
\begin{equation}
\langle \hat a_\phi \rangle = \frac{\langle \hat
S_{-\phi - \pi / 2} \rangle_0}{\sqrt{2 | \langle \hat S_z \rangle_0 |}}.
\end{equation}

\begin{figure}[tb]
\leavevmode\epsfysize=74mm \epsfbox{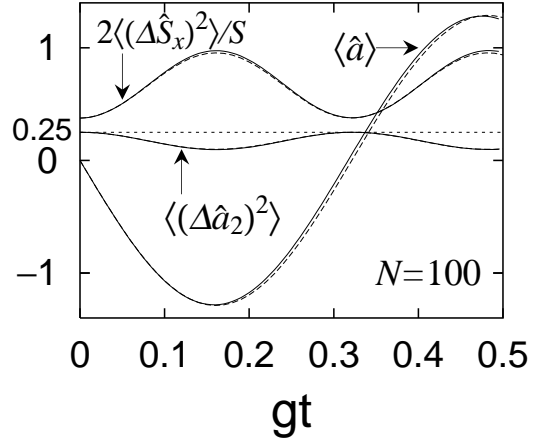}
\caption{
Time evolutions of the normalized variance $2 \langle (\Delta \hat S_x)^2
\rangle / S$ of the atoms, and the amplitude $\langle \hat a \rangle$ and
the variance $\langle (\Delta \hat a_2)^2 \rangle$ of the photon field
emitted from them.
The atomic state at $gt = 0.14$ in Fig.~\protect\ref{f:compare} is used
as the initial atom state.
The field is initially in the vacuum state.
The solid curves show the numerical solutions and the dashed curves show
the approximate solutions (\protect\ref{ampsol}a) and
(\protect\ref{flucsol}).
}
\label{f:evolution2}
\end{figure}

Figure \ref{f:evolution2} shows time evolutions of the amplitude and the
variance of the photon field, where the initial atomic state is the SAS of
100 atoms.
This atomic state is prepared by the method discussed in
Sec.~\ref{s:generation} (the state at $gt = 0.14$ in
Fig.~\ref{f:compare}).
Since the tilting angle $\tan^{-1}(-\langle \hat S_y \rangle_0 / \langle
\hat S_z \rangle_0) = 0.258$ is small, the small-angle approximation is
valid.
The solutions (\ref{ampsol}) and (\ref{flucsol}) are used for the
theoretical curves in Fig.~\ref{f:evolution2} (dashed curves).
One can see that the analytic results agree well with the numerical ones
(solid curves), and $\langle (\Delta \hat a_2)^2 \rangle$ goes below the
SQL of $1 / 4$.
It can be shown numerically that the second and the later minimums of
$\langle (\Delta \hat a_2)^2 \rangle$ are larger than the first minimum,
and hence we should switch off the interaction when the first minimum is
reached.

\subsection{Tailor-made radiation from squeezed atoms}

\begin{figure}[tb]
\leavevmode\epsfysize=148mm \epsfbox{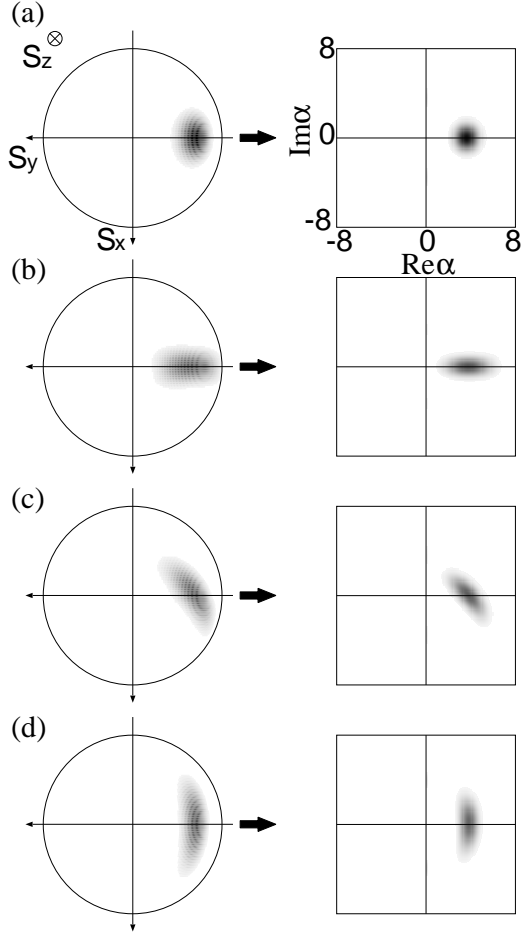}
\caption{
The quasi-probability distributions of 100 atoms (left) and those of the
photons emitted from the atoms (right).
In (a) the atoms are prepared in a coherent atomic state, and in (b), (c),
and (d), they are prepared in squeezed atom states.
In (b), (c), and (d) the uncertainty ellipses are turned around by angles
0, $\pi / 4$, $\pi / 2$, respectively.
The mean spin vectors are tilted by $\pi / 4$ from the negative $S_z$
axis.
The spin sphere is seen from the negative $S_z$ axis.
}
\label{f:radiation}
\end{figure}

As illustrated in Fig.~\ref{f:ellipses}, the quasi-probability
distribution of the emitted photon state is like a projection from that of
the prepared atomic state.
This observation, together with the solutions (\ref{ampsol}) and
(\ref{flucsol}), suggests to us that we can manipulate the direction of
displacement and the direction of squeezing of photons by controlling the
spin vector of the SAS.
The rotation of the spin vector about an axis on the $S_x$-$S_y$ plane can
be made by applying maser or laser with classical intensity to the atoms
as described by the Hamiltonian (\ref{Hcl}).
The rotation about the $S_z$ axis is realized by applying a dc magnetic
field which causes a temporal detuning by the Zeeman shift.
Combining these two processes, we can manipulate both the spin vector and
the direction of squeezing.
By manipulating the SAS in the spin space, we can control the uncertainty
ellipse of the photon field on the complex-$\alpha$ plane.
Figure \ref{f:radiation} shows the quasi-probability distributions of 100
atoms (left panels) and those of the emitted photon states (right panels).
In Fig.~\ref{f:radiation}(a) the CAS is used, and in
Figs.~\ref{f:radiation}(b)-(d) the atom states are prepared in the SASs
by the method discussed in Sec.~\ref{s:generation}, where the parameters
are optimized to obtain the maximum degree of spin squeezing ($\alpha =
6.8$, $gt = 0.19$).
The tilting angle of the spin vector from the negative $S_z$ axis is taken
to be $\pi / 4$ in Figs.~\ref{f:radiation}(a)-(d), and the uncertainty
ellipses are turned around by 0, $\pi / 4$, and $\pi / 2$ in
Figs.~\ref{f:radiation}(b), (c), and (d), respectively.
One finds that the fluctuation profiles of the atomic states are rather
faithfully transferred to those of the emitted photon states.
Figures \ref{f:radiation}(c) and (d) suggest that not only amplitudes and
fluctuations but also higher-order moments of atom states are transferred
to those of the photon states.
We have thus demonstrated that by manipulating the SAS, we can control
quantum statistical properties of the photon field at our disposal, which
we would like to call {\it tailor-made radiation}.

The squeezing of photons in the direction of phase can be obtained only if
the atomic state is squeezed in the azimuth direction as in
Fig.~\ref{f:radiation}(b).
Although the CAS can produce the photon-number squeezed
state~\cite{Heidmann} as in Fig.~\ref{f:radiation}(a), where the Fano
factor is 0.81, it never produces the phase-squeezed photon state by any
rotation on the spin sphere.
This can be verified numerically, and can also be deduced from the fact
that the projection of the fluctuation profile on the complex-$\alpha$
plane from the spin sphere can never be squeezed in the direction of the
phase if the fluctuation profile on the spin sphere is isotropic.
To produce not only the amplitude-squeezed state but also the
phase-squeezed state, the atom state must therefore be squeezed in the
sense of the definition (\ref{SSSdef}).

\subsection{Available range of the tailor-made radiation}

Let us discuss the range of photon squeezing that is available by our
method.
We use the SAS generated by the interaction between the totally excited
atoms and the coherent state of the photon field with an optimum amplitude
as discussed in Sec.~\ref{s:generation}.
The available range of the emitted photon field is obtained by plotting
time evolutions of the radiation processes for various initial tilting
angles of the spin vector of the SAS.

\begin{figure}[tb]
\leavevmode\epsfysize=122mm \epsfbox{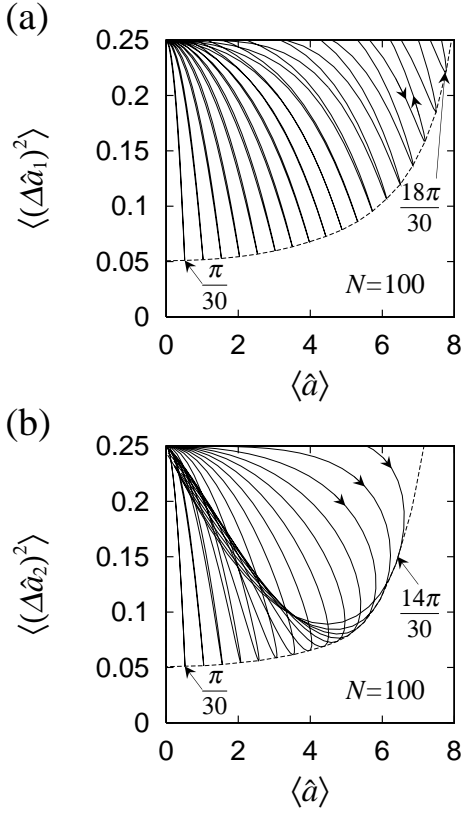}
\caption{
Time evolution of the amplitudes and the variances of the quadrature
components of the photon state emitted from the squeezed 100 atoms.
Each trajectory is drawn with the initial tilting angle of the mean spin
vector at every $\pi / 30$.
The squeezed atom states are prepared in the same manner as in
Fig.~\protect\ref{f:radiation} and rotated to the states which are
squeezed in the latitudinal direction in (a) (as in
Fig.~\protect\ref{f:radiation}(d)), and in the longitudinal direction in
(b) (as in Fig.~\protect\ref{f:radiation}(b)).
The dashed curves delimit the regions that the trajectories can reach.
}
\label{f:quadregion}
\end{figure}

Figure \ref{f:quadregion} shows time evolutions of the amplitudes and the
variances of the quadrature amplitudes of the photon states emitted from
the SASs of 100 atoms.
Each trajectory is drawn with the initial tilting angle of the mean spin
vector at every $\pi / 30$.
In Fig.~\ref{f:quadregion}(a), the SASs are prepared in the states
squeezed in the longitudinal direction, as in Fig.~\ref{f:radiation}(b).
The emitted photon states are therefore out-of-phase squeezed states.
In Fig.~\ref{f:quadregion}(b), the initial SASs are squeezed in the
latitudinal direction as in Fig.~\ref{f:radiation}(d), and the emitted
photon states are therefore in-phase squeezed states.
We find that in Fig.~\ref{f:quadregion}(a) the trajectories tend to return
the same paths, whereas in Fig.~\ref{f:quadregion}(b) the trajectories
tend to round downward.
This indicates that in the case of in-phase squeezing the energy exchange
and the fluctuation exchange between the atoms and the photon field tend
to occur synchronously, and in the out-of-phase squeezing the fluctuation
exchange tends to be delayed against the energy exchange.
When we draw the overlap region of Figs.~\ref{f:quadregion}(a) and (b), we 
can obtain the available range of the quadrature-amplitude squeezed
state.
It can be shown that the larger number of atoms can produce the wider
range of $|\langle \hat a \rangle|$ and $\langle (\Delta \hat a_\phi)^2
\rangle$~\cite{Saito}.
This is due to the fact that the larger is the number of atoms the larger
will be the degree of squeezing of the SAS, as shown in
Fig.~\ref{f:dependence}.

\begin{figure}[tb]
\leavevmode\epsfysize=74mm \epsfbox{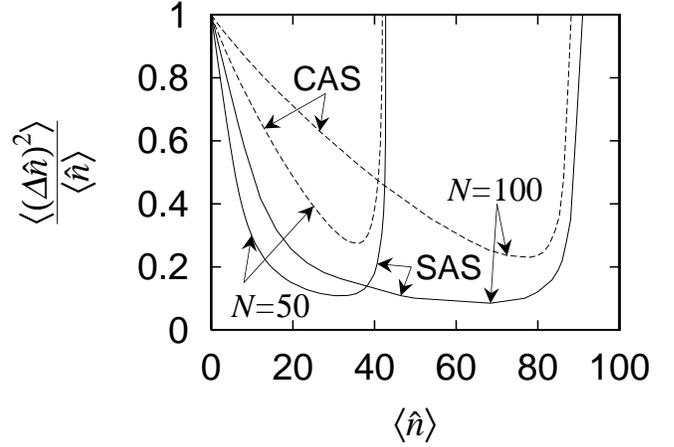}
\caption{
Ranges of the average photon number $\langle \hat n \rangle$ and the Fano
factor $\langle (\Delta \hat n)^2 \rangle / \langle \hat n \rangle$ of the
photon field that can be obtained by the squeezed atom state (SAS) and the
coherent atom state (CAS) of 50 and 100 atoms.
The SASs are prepared in the same manner as in
Fig.~\protect\ref{f:radiation}.
The regions above the curves show the available photon states.
The solid curves show the results of the SASs and the dashed ones show
those of the CASs.
}
\label{f:numregion}
\end{figure}

The ranges of the average photon number $\langle \hat n \rangle$ and the
Fano factor $\langle (\Delta \hat n)^2 \rangle / \langle \hat n \rangle$
available from the SASs and the CASs of 50 and 100 atoms are shown in
Fig.~\ref{f:numregion}.
It is found that for a given number of atoms the SAS can suppress the
photon-number fluctuation more effectively than the CAS.
The range of 100 atoms does not cover that of 50 atoms in
Fig.~\ref{f:numregion}.
The SAS of 50 atoms can produce photon states having smaller Fano factors
than the SAS of 100 atoms when the average photon number is less than
about 40.
For a given average photon number, therefore, there is an optimal
number of atoms to produce the best photon-number squeezed state.

\begin{figure}[tb]
\leavevmode\epsfysize=74mm \epsfbox{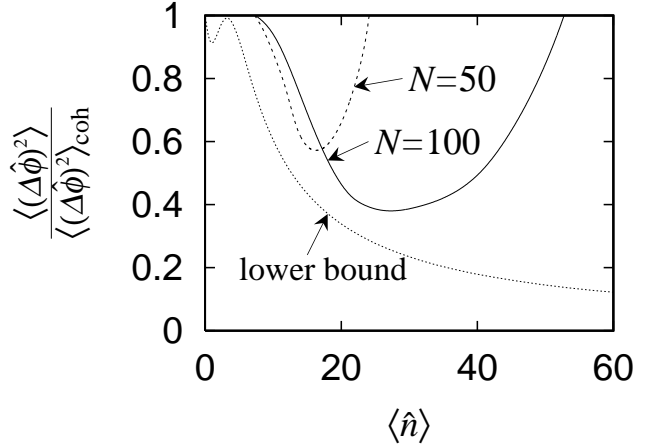}
\caption{
Available ranges of the average photon number $\langle \hat n \rangle$ and
the phase fluctuation $\langle (\Delta \hat\phi)^2 \rangle$ normalized by
that of the coherent state having the same average photon number $\langle
(\Delta \hat\phi)^2 \rangle_{\rm coh}$.
The regions above the curves show the photon states that can be obtained
by our method.
The squeezed atom states of 50 and 100 atoms are prepared in the same
manner as in Fig.~\protect\ref{f:radiation}.
The dotted curve shows the lower bound of $\langle (\Delta \hat\phi)^2
\rangle / \langle (\Delta \hat\phi)^2 \rangle_{\rm coh}$ of the photon
field.
}
\label{f:phaseregion}
\end{figure}

The ranges of the average photon number $\langle \hat n \rangle$ and the
phase fluctuation $\langle (\Delta \hat \phi)^2 \rangle$ available from
the SASs of 50 and 100 atoms are shown in Fig.~\ref{f:phaseregion}.
Here we use the phase operator proposed by Pegg and Barnett~\cite{PB}.
When $\langle \hat a \rangle$ is real and positive, the variance of the
phase is expressed as
\begin{equation}
\langle (\Delta \hat\phi)^2 \rangle = \frac{\pi^2}{3} + \sum_{n \neq n'}
\frac{2 (-1)^{n - n'}}{(n - n')^2} \;\; {}_F \langle n' | \hat \rho_F | n
\rangle_F,
\end{equation}
where $\hat \rho_F$ is the density operator of the photon field and $|n
\rangle_F$ is the photon-number state.
Figure \ref{f:phaseregion} shows the variance of the phase $\langle (\Delta
\hat\phi)^2 \rangle$ normalized by that of the coherent state having the
same average photon number $\langle (\Delta \hat\phi)^2 \rangle_{\rm
coh}$.
Here the phase is defined as squeezed when $\langle (\Delta \hat\phi)^2
\rangle / \langle (\Delta \hat\phi)^2 \rangle_{\rm coh}$ is below unity.
The dotted curve in Fig.~\ref{f:phaseregion} shows minimum values of
$\langle (\Delta \hat\phi)^2 \rangle / \langle (\Delta \hat\phi)^2
\rangle_{\rm coh}$ for given average photon numbers, which are obtained by 
the method of Lagrange multipliers~\cite{DAriano} (see
appendix~\ref{app:Lagrange}).
The range of 100 atoms does not completely include that of 50 atoms as in
the case of the Fano factor, which indicates that for a given average
photon number there is an optimal number of atoms to reduce the phase
fluctuation.

\begin{figure}[tb]
\leavevmode\epsfysize=125mm \epsfbox{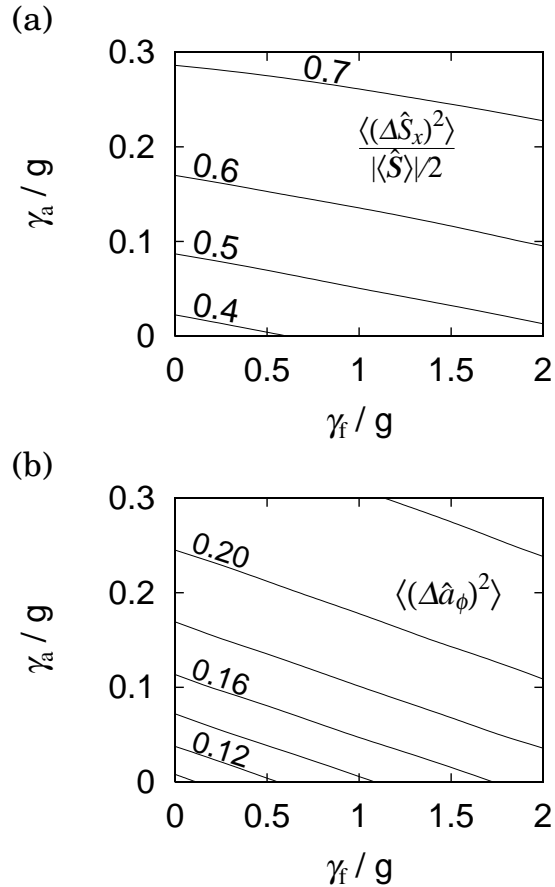}
\caption{
(a) The contour plot of the minimum values of the squeezing factor $2
\langle (\Delta \hat S_x)^2 \rangle / |\langle \hat{\bf S} \rangle|$ of
the SASs obtained by the interaction of 10 atoms with the coherent state
of the photon field.
The amplitude of the coherent state is optimized to obtain the maximum
degree of squeezing for each $\gamma_f$ and $\gamma_a$.
(b) The contour plot of the minimum values of $\langle (\Delta \hat
a_\phi)^2 \rangle$ of the photon field emitted from the squeezed atoms
prepared in (a). 
}
\label{f:dissip}
\end{figure}

In experimental situations, loss of photons in the cavity and spontaneous
emission of atoms are unavoidable, and we therefore evaluate how much
cavity loss and spontaneous emission are allowed in order not to destroy
the squeezing of the atoms and that of the photon field.
We adopt the master-equation approach to take into account the
effects of dissipation.
The master equation in the presence of cavity loss and spontaneous
emission is given by~\cite{Louisell}
\begin{eqnarray} \label{master}
\frac{\partial \hat\rho}{\partial t} & = & \frac{i}{\hbar} [\hat\rho, \hat
H^{\rm rot}] + \frac{\gamma_f}{2} \left( 2 \hat a \hat \rho \hat a^\dagger
- \hat a^\dagger \hat a \hat \rho - \hat \rho \hat a^\dagger \hat a
\right) \nonumber \\
& & + \frac{\gamma_a}{2} \left( 2 \hat S_- \hat \rho \hat S_+ - \hat
S_+ \hat S_- \hat \rho - \hat \rho \hat S_+ \hat S_- \right),
\end{eqnarray}
where $\hat \rho$ denotes the density operator of both the atoms and the
photon field, and $\gamma_f^{-1}$ and $\gamma_a^{-1}$ are the lifetimes of
a single photon and a single atom in the cavity.
We obtain time evolution of the density operator by numerically
integrating the master equation (\ref{master}) by the Runge-Kutta method.
Figure~\ref{f:dissip}(a) shows the contour plot of the minimum attainable
values of the squeezing factor $2 \langle (\Delta \hat S_x)^2 \rangle /
|\langle \hat{{\bf S}} \rangle |$ of the SASs obtained by the interaction
of 10 atoms with the coherent state of the photon field.
The amplitude of the coherent state is optimized to obtain the maximum
degree of squeezing for each $\gamma_f$ and $\gamma_a$.
Figure~\ref{f:dissip}(b) shows the contour plot of the minimum values of
$\langle (\Delta \hat a_\phi)^2 \rangle$ of the photon field emitted from
the squeezed atoms prepared in Fig.~\ref{f:dissip}(a).
The parameters $\gamma_f$ and $\gamma_a$ in the radiation process are
assumed to have the same values as in the preparation of the SAS.
These results show that the generation of the SAS and the squeezed
radiation are possible even in the presence of dissipation in
experimentally feasible situations.
We will discuss some concrete numbers in the next section.

\section{Possible Experimental Situations}
\label{s:exp}

We discuss possible experimental situations to implement our theory.
Our procedure of generating quantum-controlled few-photon states consists
of three stages: (1) preparation of the SAS, (2) manipulation of the SAS
(rotation of the spin vector in the spin space), and (3) radiation from
these atoms.

\begin{figure}[tb]
\leavevmode\epsfysize=90mm \epsfbox{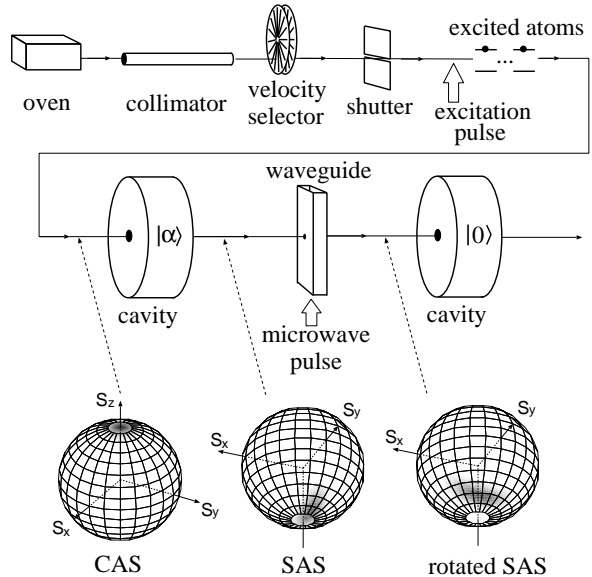}
\caption{
Schematic illustration of an experimental setup to implement the
tailor-made radiation.
The state of the atoms at each stage is shown with the spin
quasi-probability distribution.
A bunch of two-level excited atoms that is in a coherent atom state (CAS)
is prepared by an oven, a collimator, a velocity selector, a shutter, and
a pulse that excites the atoms.
The atoms then go into the first cavity and interact with a coherent state
of the photon field $|\alpha\rangle$.
The output atoms are in a squeezed atom state (SAS).
By the interaction with a microwave pulse in a waveguide, the mean spin
vector is rotated to a desired direction.
The atoms finally go into the second cavity and emit photons there.
Left in the second cavity is the desired few-photon state.
}
\label{f:setup}
\end{figure}

A simplest realization of our theory would be to fly a bunch of atoms
through two cavities and a waveguide as schematically illustrated in
Fig.~\ref{f:setup}.
This type of experiment may be done in a microwave regime, since the atoms 
are required to be within a region much smaller than the wavelength.
If we use, for example, the $63{\rm p}_{3/2} \leftrightarrow 61{\rm
d}_{3/2}$ transition of rubidium atoms, the resonant frequency is 21.5
GHz, the wavelength is $\lambda \sim 10^{-2}$ m, and the coupling
constant is $g \sim 10^4$ Hz.
First, an atomic beam from an oven is collimated and velocity-selected.
The variance of the velocity of the atoms must be $\Delta v \ll \lambda /
T \sim 10^{2}$ m/s, where $T$ is the time it takes the atoms to pass
through the apparatus.
A mechanical shutter can prepare a bunch of atoms from the atomic beam.
The atoms in the bunch are then excited to the Rydberg state that is the
upper state of the relevant two-levels, and enter the first cavity in
which the photon field is prepared to be in a coherent state $|\alpha
\rangle$.
The SAS is generated there by the higher-order interaction of the atoms
with the coherent state.
The interaction time is $gt_1 \sim 10^{-1}$, i.e., $t_1 \sim 10^{-5} {\rm
s}$, e.g., in the situation in Fig.~\ref{f:evolution1}.
The velocity of the atoms is therefore required to be $v \sim 10^3 {\rm
m/s}$.
The atoms then pass through a waveguide, where the atoms are irradiated
by a pulse of microwave with classical intensity, by which the spin vector
representing the state of collective atoms is rotated.
To control the rotation axis of the spin vector, the relative phase
between the microwave and the coherent state in the first cavity must be
controlled.
The irradiation time of the classical field is much shorter than the
interaction time in both cavities.
Finally, the atoms pass through the second cavity in which the desired
state of photons is emitted from the atoms.
The interaction time is $gt_2 \sim 10^{-1}$, i.e., $t_2 \sim 10^{-5} {\rm
s}$, e.g., in the situation in Fig.~\ref{f:evolution2}.
The atoms thus pass through the two cavities within a few periods of
duration $10^{-5} {\rm s}$, which is much shorter than the lifetime of the
Rydberg atoms $\sim 10^{-3} {\rm s}$ and the cavity lifetime $\sim 10^{-1}
{\rm s}$~\cite{Rempe}.
From Fig.~\ref{f:dissip}, this cavity lifetime corresponding to $\gamma_f
/ g \sim 10^{-3}$ does not affect the squeezing.
If the circular Rydberg states are used, the lifetime is $\sim 1$ s, and
decays from the relevant levels become negligible.
Since the microwave frequency is used, the temperature should be lower
than $\sim 1$ K in order to make the average number of thermal photons in
the cavity much smaller than that of the produced photons.

Another possible scheme is to use atoms confined in an ion trap or a
magnetic trap in which the quantized kinetic motion of the atoms replaces
the role of photons.
Wineland {\it et al.}~\cite{Wineland} proposed the JC interaction between
the Zeeman doublet of electronic states of each ion and the
center-of-mass (CM) motion of an ensemble of ions via the inhomogeneous
magnetic field.
They pointed out that the stimulated Raman transition can also be
used to couple the internal states of each ion to the CM motion of
ions~\cite{Wineland,Heinzen}.
In these models the operators $\hat a$ and $\hat a^\dagger$ in the JC
Hamiltonian (\ref{JCH}) are not for photons but for the quantized CM
motion of ions in a harmonic trap.
By using the stimulated Raman technique, our theory might be implemented
as follows.
First, the internal levels of the trapped ions are excited and the CM
motion is cooled to the ground state~\cite{Diedrich}.
The CM motions of two orthogonal directions, say the $z$ and $x$
directions, correspond to the photon fields in the first and the second
cavities in the method discussed in the preceding paragraph.
In the first stage, the coherent state of the CM motion in the $z$
direction is prepared and the Raman beams in this direction are applied.
The coherent state of the CM motion can be generated by sudden
displacement of the trap center.
When the atomic internal state becomes the SAS, the Raman beams are
switched off.
In the second stage, the Raman beams that do not affect the CM motion are
applied, which rotate the spin vector in the spin space.
In the third stage, the Raman beams in the $x$ direction are applied, and
the internal states of the ions are coupled to the CM motion in the
$x$ direction.
By this coupling the information of the internal states is transferred to 
the CM motion in the $x$ direction, which may be called a tailor-made
motional state.
Although this is not radiation, the method using the trapped atoms might
be used to test our theory.

The use of dielectric spheres as optical cavities might be another
possibility, where the optical whispering gallery (WG) mode in the
microsphere is employed.
With the microsphere cavity, very low threshold lasing has been
observed~\cite{Tseng,Gonokami}, and the Q value of more than $10^9$ has
been achieved with highly transparent silica glass optical-fiber
material~\cite{Collot}.
The atoms are fixed on the substrate and are coupled to evanescent waves
of two microspheres which are placed very closely.
They have the slightly different resonant frequencies $\omega_1$ and
$\omega_2$.
The optical WG mode in the first microsphere is prepared in a coherent
state, while that in the second one is prepared in the vacuum state.
In the first stage, the atoms are brought into resonance
with the WG mode frequency in the first microsphere $\omega_1$,
and are far from resonant with that of the second microsphere $\omega_2$.
This can be done by Zeeman-shifting the transition frequency of the atoms
by a magnetic field.
When the atomic state becomes the SAS by interacting with the coherent
state, the interaction with the WG mode in the first microsphere is
switched off by switching off the magnetic field.
In the second stage, the spin vector is rotated by applying a laser pulse
resonant with the transition frequency of the atoms.
In the third stage, by applying an appropriate magnetic field,
the atoms are brought into resonance with the WG mode of the second
microsphere $\omega_2$.
By switching off the magnetic field, the desired photon state is left
in the second microsphere.
The coupling constant $g$ can be of order $10^8$, and $\gamma_f / g \sim
10^{-2}$ and $\gamma_a / g \sim 10^{-1}$, where the spontaneous emission
rate of an atom in the free space is assumed.
From Fig.~\ref{f:dissip}, we find that both the SAS and the squeezed
photon state are not washed out by the effects of dissipation.

\section{Conclusions}
\label{s:conclusions}

In conclusion, we have shown that quantum fluctuations of few-photon
states can be controlled by using the SAS.
This controllability is based on the fact that quantum fluctuations of the
atoms are faithfully transferred to those of the emitted photons.
The correspondence shown in Fig.~\ref{f:radiation} between the
quasi-probability distribution on the spin sphere and that on the
complex-$\alpha$ plane indicates that a variety of photon states can be
produced by merely rotating the spin vector of the SAS.
We also found that this manipulation of few-photon states is possible only 
if the atoms are in the SAS.
Although the CAS can produce the photon-number squeezed state,
the degree of squeezing is lower than that of the photon state produced by
the SAS, and the phase squeezed state can never be produced by the CAS.
The possible experimental situations to implement our theory were
discussed.
By these schemes, we can generate the quantum-controlled few-photon state
in the microcavity, and the quantum-controlled center-of-mass motion of
trapped atoms.

\acknowledgements

One of the authors (H.S.) acknowledges support by a Research Fellowship of
the Japan Society for the Promotion of Science for Young Scientists.

\appendix

\section{Vanishing expectation values in the Jaynes-Cummings interaction}
\label{s:vanish}

It is assumed in Sec.~\ref{s:generation} that the expectation values
$\langle \hat S_x \rangle$ and $\langle \hat a_2 \rangle$ always vanish if
the time evolution is governed by the Hamiltonian (\ref{H}) and when the
initial state is $|\alpha \rangle \otimes |S, M = S \rangle$ with real
$\alpha$.
In this appendix we give a general condition for this to be true.

Since an expectation value of an Hermitian operator, say $\hat{\cal O}$,
is real, it follows that
\begin{eqnarray} \label{b1}
\langle e^{i\hat H^{\rm rot} t} \hat{\cal O} e^{-i\hat H^{\rm rot} t}
\rangle_0 & = & \langle e^{i\hat H^{\rm rot} t} \hat{\cal O} e^{-i\hat
H^{\rm rot} t} \rangle_0^* \nonumber \\
& = & \langle e^{-i\hat H^{\rm rot} t} \hat{\cal O}^* e^{i\hat H^{\rm rot}
t} \rangle_0,
\end{eqnarray}
where the expectation values are taken with respect to the initial state 
$|\alpha \rangle \otimes |S, M = S \rangle$, and $\hat{\cal O}^*$
denotes an operator whose matrix elements are complex conjugates of those
of $\hat{\cal O}$.
In the second line of Eq.~(\ref{b1}) we used the fact that the matrix
element of the Hamiltonian (\ref{H}), 
\begin{eqnarray}
& & \langle n | \langle S, M | \hat H^{\rm rot} | S, M' \rangle | n'
\rangle \nonumber \\
& = & g\hbar \Bigl[ \sqrt{n + 1} \sqrt{(S + M) (S - M + 1)} \delta_{n, n'
- 1} \delta_{M, M' + 1} \nonumber \\
& & + \sqrt{n} \sqrt{(S - M) (S + M + 1)} \delta_{n, n' + 1} \delta_{M, M'
- 1} \Bigr],
\end{eqnarray}
is real and hence $\hat H^{{\rm rot} *} = \hat H^{\rm rot}$.
By a unitary transformation $e^{i\pi \hat S_z}$ we have $e^{i\pi \hat S_z} 
\hat H^{\rm rot} e^{-i\pi \hat S_z} = -\hat H^{\rm rot}$ and $e^{i\pi \hat
S_z} |\alpha \rangle \otimes |S, M = S \rangle = e^{i\pi S} |\alpha
\rangle \otimes |S, M = S \rangle$.
Applying this unitary transformation to the second line of (\ref{b1}),
the expectation value becomes
\begin{equation} \label{b2}
\langle e^{i\hat H^{\rm rot} t} \hat{\cal O} e^{-i\hat H^{\rm rot} t}
\rangle_0 = \langle e^{i\hat H^{\rm rot} t} e^{i\pi \hat S_z} \hat{\cal
O}^* e^{-i\pi \hat S_z} e^{-i\hat H^{\rm rot} t} \rangle_0.
\end{equation}
Therefore, if $e^{i\pi \hat S_z} \hat{\cal O}^* e^{-i\pi \hat S_z} =
-\hat{\cal O}$, the expectation value (\ref{b2}) must vanish.
The operators $\hat S_x$ and $\hat a_2$ meet this condition.
In general, an expectation value of an Hermitian operator that consists of 
operator products in which $\hat S_x$ and $\hat a_2$ appear odd-numbered
times always vanishes.
General conditions required for the initial state of the photon field
$\hat \rho_F$ and that of the atoms $\hat \rho_A$ are $\hat \rho_F^* =
\hat \rho_F$ and $e^{i\pi \hat S_z} \hat \rho_A e^{-i\pi \hat S_z} = \hat
\rho_A$.

\section{Derivation of the approximate solutions
(\protect\ref{SOLUTIONS})}
\label{s:solution}

In this appendix, we derive the approximate solutions (\ref{SOLUTIONS}).
It is convenient to define~\cite{Heidmann}
\begin{mathletters}
\begin{eqnarray}
\hat a_i' & \equiv & \frac{\hat a_i}{\sqrt{N}} \;\;\; (i = 1, 2), \\
\hat S_\mu' & \equiv & \frac{\hat S_\mu}{N} \;\;\; (\mu = x, y, z), \\
\tau & \equiv & g \sqrt{N} t,
\end{eqnarray}
\end{mathletters}
in order to estimate errors of the approximation.
The equations of motion for these operators have the forms
\begin{mathletters} \label{Heq}
\begin{eqnarray}
\partial_\tau \hat S_x' & = & -2 \hat a_2' \hat S_z', \\
\partial_\tau \hat S_y' & = & -2 \hat a_1' \hat S_z', \\
\partial_\tau \hat S_z' & = & 2 (\hat a_1' \hat S_y' + \hat a_2' \hat S_x'), \\
\partial_\tau \hat a_1' & = & -\hat S_y', \\
\partial_\tau \hat a_2' & = & -\hat S_x'.
\end{eqnarray}
\end{mathletters}
We assume that the initial state is the CAS $|\theta = 0, \phi \rangle$
for the atoms and the coherent state $|\alpha \rangle$ for the photon
field, where $\alpha$ is taken to be real without loss of generality.
Taking the expectation values of Eqs.~(\ref{Heq}) yields
\begin{mathletters} \label{avgeom}
\begin{eqnarray}
\partial_\tau \langle \hat S_y' \rangle & = & -2 \langle \hat a_1' \rangle 
\langle \hat S_z' \rangle - 2 \langle \Delta \hat a_1' \Delta \hat S_z'
\rangle, \\
\partial_\tau \langle \hat S_z' \rangle & = & 2 \langle \hat a_1' \rangle 
\langle \hat S_y' \rangle - 2 \langle \Delta \hat a_1' \Delta \hat S_y'
\rangle, \\
\partial_\tau \langle \hat a_1' \rangle & = & -\langle \hat S_y' \rangle,
\end{eqnarray}
\end{mathletters}
where $\Delta \hat{\cal O} \equiv \hat{\cal O} - \langle \hat{\cal O}
\rangle$ for any operators.
It can be shown that $\langle \hat S_y' \rangle$, $\langle \hat S_z'
\rangle$, and $\langle \hat a_1' \rangle$ are of order unity, and $\langle 
\Delta \hat a_1' \Delta \hat S_z' \rangle$ and $\langle \Delta \hat a_1'
\Delta \hat S_y' \rangle$ are of order $1 / N$.
If we neglect relative errors of $1 / N$, the second terms of
Eqs.~(\ref{avgeom}a) and (\ref{avgeom}b) can be neglected, giving
\begin{mathletters} \label{avgeom2}
\begin{eqnarray}
\partial_\tau \langle \hat S_y' \rangle & = & -2 \langle \hat a_1' \rangle 
\langle \hat S_z' \rangle, \\
\partial_\tau \langle \hat S_z' \rangle & = & 2 \langle \hat a_1' \rangle 
\langle \hat S_y' \rangle, \\
\partial_\tau \langle \hat a_1' \rangle & = & -\langle \hat S_y' \rangle.
\end{eqnarray}
\end{mathletters}
If we set
\begin{mathletters} \label{pendulum}
\begin{eqnarray}
\langle \hat S_y' \rangle & = & \frac{1}{2} \sin\theta(\tau), \\
\langle \hat S_z' \rangle & = & \frac{1}{2} \cos\theta(\tau), \\
\langle \hat a_1' \rangle & = & -\frac{1}{2} \partial_\tau \theta(\tau),
\end{eqnarray}
\end{mathletters}
the equations of motion (\ref{avgeom}) reduce to
\begin{equation}
\partial_\tau^2 \theta(\tau) = \sin\theta(\tau),
\end{equation}
which has the same form as the equation of motion for the mechanical
pendulum.
The angular velocity of the pendulum corresponds to the field amplitude.
The solutions for the initial condition $\theta(0) = 0$ and
$\partial_\tau \theta(0) = -2 \alpha'$, where $\alpha' \equiv \alpha /
\sqrt{N}$, can be expressed in terms of Jacobi's elliptic functions as
\begin{mathletters} \label{sol1}
\begin{eqnarray}
\langle \hat S_y' \rangle & = & -\sqrt{1 - m} \; {\rm sd}(u|m) {\rm
cd}(u|m), \\
\langle \hat S_z' \rangle & = & \frac{1}{2} \left[ 2 {\rm cd}^2(u|m) - 1
\right], \\
\langle \hat a_1' \rangle & = & \alpha' \; {\rm nd}(u|m), \\
\langle \hat a_1' \rangle & = & \langle \hat S_x' \rangle = 0,
\end{eqnarray}
\end{mathletters}
where $u \equiv \tau \sqrt{1 + \alpha'^2}$ and $m \equiv 1 / (1 +
\alpha'^2)$.

The equations of motion for variances are written as
\begin{mathletters} \label{fleq}
\begin{eqnarray}
\partial_\tau \langle (\Delta \hat a_2')^2 \rangle & = & -2 \langle \Delta
\hat a_2' \Delta \hat S_x' \rangle, \\
\partial_\tau \langle (\Delta \hat S_x')^2 \rangle & = & -4 \langle \Delta
\hat a_2' \Delta \hat S_x' \rangle \langle \hat S_z' \rangle + 2 \langle
\Delta \hat S_x' \Delta \hat S_z' \Delta \hat a_2' \rangle \nonumber \\
& & + 2 \langle
\Delta \hat S_z' \Delta \hat S_x' \Delta \hat a_2' \rangle, \\
\partial_\tau \langle \Delta \hat a_2' \Delta \hat S_x' \rangle & = &
-\langle (\Delta \hat S_x')^2 \rangle - \langle (\Delta \hat a_2')^2
\rangle \langle \hat S_z' \rangle \nonumber \\
& & - \langle (\Delta \hat a_2')^2 \Delta \hat S_z' \rangle.
\end{eqnarray}
\end{mathletters}
It can be shown that the second-order fluctuations, such as $\langle
(\Delta \hat S_x')^2 \rangle$, are of order $1 / N$, and that the
third-order fluctuations, such as $\langle \Delta \hat S_x' \Delta \hat
S_z' \Delta \hat a_2' \rangle$, are of order $1 / N^2$.
Neglecting the third-order fluctuations in Eqs.~(\ref{fleq}), we have
\begin{mathletters} \label{fleq2}
\begin{eqnarray}
\partial_\tau \langle (\Delta \hat a_2')^2 \rangle & = & -2 \langle \Delta
\hat a_2' \Delta \hat S_x' \rangle, \\
\partial_\tau \langle (\Delta \hat S_x')^2 \rangle & = & -4 \langle \Delta
\hat a_2' \Delta \hat S_x' \rangle \langle \hat S_z' \rangle, \\
\partial_\tau \langle \Delta \hat a_2' \Delta \hat S_x' \rangle & = &
-\langle (\Delta \hat S_x')^2 \rangle - \langle (\Delta \hat a_2')^2
\rangle \langle \hat S_z' \rangle.
\end{eqnarray}
\end{mathletters}
Using the form of $\langle \hat S_z' \rangle$ in Eq.~(\ref{sol1}b), which
has at most a relative error of $1 / N$, Eqs.~(\ref{fleq}) reduce to the
closed differential equations with relative errors $1 / N$.
They have three independent sets of solutions, and two of them are
obtained as
\begin{eqnarray}
& & \left( \begin{array}{c} \langle (\Delta \hat a_2')^2 \rangle \\ \langle
(\Delta \hat S_x')^2 \rangle \\ \langle \Delta \hat a_2' \Delta \hat S_x'
\rangle \end{array} \right) = \left( \begin{array}{l} \frac{1}{N} {\rm 
nd}^2(u|m) \\ \frac{m}{N} {\rm sd}^2(u|m) {\rm cd}^2(u|m) \\
-\frac{\sqrt{m}}{N} {\rm sd}(u|m) {\rm cd}(u|m) {\rm nd}(u|m) \end{array}
\right), \nonumber \\
& & \left( \begin{array}{l} \frac{1}{N} {\rm nd}^2(u|m) E^2(u|m) \\
\frac{m}{N} \left[ m \; {\rm sd}(u|m) {\rm cd}(u|m) E(u|m) + {\rm dn}(u|m) 
\right]^2 \\ \scriptstyle -\frac{\sqrt{m}}{N} \left[ m \; {\rm sd}(u|m)
{\rm cd}(u|m) E(u|m) + {\rm dn}(u|m) \right] {\rm nd}(u|m) E(u|m)
\end{array} \right). \nonumber \\
& & 
\end{eqnarray}
The linear combination of these solutions to satisfy the initial
conditions $\langle (\Delta \hat S_x')^2 \rangle = \frac{1}{4N}$, $\langle
(\Delta \hat a_2')^2 \rangle = \frac{1}{4N}$, and $\langle \Delta \hat
a_2' \Delta  \hat S_x' \rangle = 0$ yields the solutions
(\ref{SOLUTIONS}).

\section{A method to minimize the phase fluctuation}
\label{app:Lagrange}

In this appendix, we briefly show a method to obtain a photon state having 
the minimum phase fluctuation, which is the dotted curve in
Fig.~\ref{f:phaseregion}.
The variance of the Pegg-Barnett phase operator of the photon state
$\sum_n c_n |n \rangle_F$ is given by
\begin{equation} \label{phivar}
\langle (\Delta \hat \phi)^2 \rangle = \frac{\pi^2}{3} + 2 \sum_{n \neq m} 
A_{nm} c_n c_m,
\end{equation}
where $A_{nm} = (-1)^{n - m} / (n - m)^2$.
The coefficients that minimize the variance (\ref{phivar}) satisfying the
constraints $\sum_n c_n^2 = 1$ and $\sum_n n c_n^2 = \bar n$ are obtained
by minimizing the function
\begin{eqnarray}
F(\{c_n\}, \lambda, \beta) & = & 2 \sum_{n \neq m} A_{nm} c_n c_m + \lambda
\left( \sum_n c_n^2 - 1 \right) \nonumber \\
& & + \beta \left( \sum_n n c_n^2 - \bar n
\right),
\end{eqnarray}
where $\lambda$ and $\beta$ are the Lagrange multipliers.
The variational problem $\partial F / \partial c_n = 0$ is equivalent to
the eigenvalue problem
\begin{equation}
\sum_{n'} \left( 2 A_{nn'} + n \beta \delta_{nn'} \right) c_{n'} + \lambda 
c_n = 0,
\end{equation}
which can be solved numerically.


\begin{references}

\bibitem{squeeze}
For collections of review articles, see, for example, R. Loudon and
P. L. Knight (eds), J. Mod. Opt. {\bf34}, (6/7) (1987);
H. J. Kimble and D. F. Walls (eds), 
J. Opt. Soc. Am. B {\bf4}, (10) (1987).

\bibitem{Kimble}
For reviews, see, for example, H. J. Kimble, Phys. Rep. {\bf 219}, 227 (1992);
in {\it Fundamental Systems in Quantum Optics} (Elsevier, Amsterdam,
1992), p. 545.

\bibitem{Slusher}
R. E. Slusher, L. W. Hollbery, B. Yurke, J. C. Mertz and J. F. Valley,
Phys. Rev. Lett. {\bf 55}, 2409 (1985).

\bibitem{Wu}
L. A. Wu, H. J. Kimble, J. L. Hall, and H. Wu, Phys. Rev. Lett. {\bf 57},
2520 (1986); E. S. Polzik, J. Carri, and H. J. Kimble,
Phys. Rev. Lett. {\bf 68}, 3020 (1992).

\bibitem{Machida87}
S. Machida, Y. Yamamoto, and Y. Itaya, Phys. Rev. Lett. {\bf 58}, 1000
(1987).

\bibitem{Tapster}
P. R. Tapster, J. G. Rarity, and J. S. Satchell, Europhys. Lett. {\bf 4},
293 (1987).

\bibitem{Hirano}
T. Hirano and T. Kuga, IEEE J. Quant. Elec. {\bf 31}, 2236 (1995).

\bibitem{Yamanishi}
M. Yamanishi, K. Watanabe, N. Jikutani, and M. Ueda, Phys. Rev. Lett.
{\bf76}, 3432 (1996).

\bibitem{Saito}
H. Saito and M. Ueda, Phys. Rev. Lett. {\bf79}, 3869 (1997).

\bibitem{Allen}
L. Allen and J. H. Eberly, {\it Optical Resonances and Two-level Atoms}
(Dover, New York, 1987).

\bibitem{Dicke}
R. H. Dicke, Phys. Rev. {\bf93}, 99 (1954).

\bibitem{fluorescence}
B. R. Mollow, in {\it Progress in Optics XIX}, edited by E. Wolf
(North-Holland, Amsterdam, 1981).

\bibitem{Meschede}
D. Meschede, H. Walther, and G. M\"uller, Phys. Rev. Lett. {\bf 54}, 551
(1985).

\bibitem{Carmichael}
H. J. Carmichael and D. F. Walls, J. Phys. B {\bf 9}, 1199 (1976).

\bibitem{Kimble77}
H. J. Kimble, M. Dagenais, and L. Mandel, Phys. Rev. Lett. {\bf 39}, 691
(1977).

\bibitem{Walls}
D. F. Walls and P. Zoller, Phys. Rev. Lett. {\bf47}, 709 (1981).

\bibitem{Arecchi}
J. M. Radcliffe, J. Phys. A {\bf4}, 313 (1971);
F. T. Arecchi, E. Courtens, R. Gilmore, and H. Thomas,
Phys. Rev. A {\bf6}, 2211 (1972).

\bibitem{Waka}
M. Ueda, T. Wakabayashi, and M. Kuwata-Gonokami, Phys. Rev. Lett.
{\bf76}, 2045 (1996).

\bibitem{Wod}
K. W\'{o}dkiewicz, Opt. Commun. {\bf51}, 198 (1984);
Phys. Rev. B {\bf32}, 4750 (1985);
K. W\'{o}dkiewicz and J. H. Eberly, J. Opt. Soc. Am. B {\bf2}, 458 (1985);
K. W\'{o}dkiewicz, P. L. Knight, S. J. Buckle, and S. M. Barnett,
Phys. Rev. A {\bf35}, 2567 (1987).

\bibitem{Macomber}
J. D. Macomber and R. Lynch, J. Chem. Phys. {\bf83}, 6514 (1985).

\bibitem{Yurke}
B. Yurke, Phys. Rev. Lett. {\bf 56}, 1515 (1986);
B. Yurke, S. L. McCall, and J. R. Klauder, Phys. Rev. A {\bf 33}, 4033
(1986).

\bibitem{Aravind}
P. K. Aravind, J. Opt. Soc. Am. B {\bf 3}, 1545 (1986).

\bibitem{Barnett}
S. M. Barnett and M. -A. Dupertuis, J. Opt. Soc. Am. B {\bf4}, 505 (1987).

\bibitem{Agarwal}
G. S. Agarwal and R. R. Puri, Phys. Rev. A {\bf 41}, 3782 (1990).

\bibitem{Kitagawa}
M. Kitagawa and M. Ueda, Phys. Rev. Lett. {\bf67}, 1852 (1991);
Phys. Rev. A {\bf47}, 5138 (1993).

\bibitem{Wineland}
D. J. Wineland, J. J. Bollinger, W. M. Itano, F. L. Moore,
and D. J. Heinzen, Phys. Rev. A{\bf46}, R6797 (1992);
D. J. Wineland, J. J. Bollinger, W. M. Itano, and D. J. Heinzen,
Phys. Rev. A{\bf50}, 67 (1994).

\bibitem{JC}
E. Jaynes and F. Cummings, Proc. IEEE {\bf51}, 89 (1963);
M. Tavis and F. W. Cummings, Phys. Rev. {\bf170}, 379 (1968);
for a recent review of the Jaynes-Cummings model, see B. W. Shore
and P. L. Knight, J. Mod. Opt. {\bf40}, 1195 (1993).

\bibitem{Kuzmich}
A. Kuzmich, K. M{\o}lmer, and E. S. Polzik, Phys. Rev. Lett. {\bf 79}, 4782
(1997).

\bibitem{MathFunc}
{\it Handbook of Mathematical Functions}, edited by M. Abramowitz and
I. A. Stegun (Dover, New York, 1969). 

\bibitem{Eberly}
J. H. Eberly, N. B. Narozhny, and J. J. Sanchez-Mondragon,
Phys. Rev. Lett. {\bf 44}, 1323 (1980).

\bibitem{Heidmann}
A. Heidmann, J. M. Raimond, and S. Reynaud, Phys. Rev. Lett. {\bf 54}, 326 
(1985).
A. Heidmann, J. M. Raimond, S. Reynaud, and N. Zagury, Opt. Commun. {\bf
54}, 189 (1985).

\bibitem{PB}
D. T. Pegg and S. M. Barnett, Europhys. Lett. {\bf 6}, 483 (1988);
J. Mod. Opt. {\bf 36}, 7 (1989).
S. M. Barnett and D. T. Pegg, Phys. Rev. A {\bf 39}, 1665 (1989).

\bibitem{DAriano}
G. M. D'Ariano and M. G. A. Paris, Phys. Rev. A {\bf 49}, 3022 (1994).

\bibitem{Louisell}
W. H. Louisell, {\it Quantum Statistical Properties of Radiation}, (Wiley,
New York, 1973).

\bibitem{Rempe}
G. Rempe, F. Schmidt-Kaler, and H. Walther, Phys. Rev. Lett. {\bf 64},
2783 (1990).

\bibitem{Heinzen}
D. J. Heinzen and D. J. Wineland, Phys. Rev. A {\bf 42}, 2977 (1990).

\bibitem{Diedrich}
F. Diedrich, J. C. Bergquist, W. M. Itano, and D. J. Wineland,
Phys. Rev. Lett. {\bf 62}, 403 (1989).

\bibitem{Tseng}
H. M. Tseng, K. F. Wall, M. B. Long, and R. K. Chang, Opt. Lett. {\bf 9},
499 (1984).

\bibitem{Gonokami}
M. Kuwata-Gonokami, K. Takeda, H. Yasuda, and K. Ema,
Jpn. J. Appl. Phys. {\bf 31}, L99 (1992).

\bibitem{Collot}
L. Collot, V. Lefevre-Sguin, M. Brune, J. M. Raimond, and S. Haroche,
Europhys. Lett. {\bf 23}, 327 (1993).

\end{references}
\end{document}